# Melting of excitonic insulator phase by an intense terahertz pulse in $Ta_2NiSe_5$

Naoki Takamura[1], Tatsuya Miyamoto[1,*], Ryohei Ikeda[1], Tetsushi Kubo[1], Masaki Yamamoto[1], Hiroki Sato[1], Yang Han[1], Takayuki Ito[1], Tetsu Sato[1], Akitoshi Nakano[2], Hiroshi Sawa[3], and Hiroshi Okamoto[1,**]

[1]*Department of Advanced Materials Science, University of Tokyo, Kashiwa, Chiba, 277-8561, Japan.*

[2]*Department of Physics, Nagoya University, 464-8602, Japan*

[3]*Department of Applied Physics, Nagoya University, Nagoya 464-8603, Japan*

\* miyamoto@k.u-tokyo.ac.jp

\*\* okamotoh@k.u-tokyo.ac.jp

**Abstract**

In this study, the optical response to a terahertz pulse was investigated in the transition metal chalcogenide $Ta_2NiSe_5$, a candidate excitonic insulator. First, by irradiating a terahertz pulse with a relatively weak electric field (0.3 MV/cm), the spectral changes in reflectivity near the absorption edge due to third-order optical nonlinearity were measured and the absorption peak characteristic of the excitonic phase just below the interband transition was identified. Next, by irradiating a strong terahertz pulse with a strong electric field of 1.65 MV/cm, the absorption of the excitonic phase was found to be reduced, and a Drude-like response appeared in the mid-infrared region. These responses can be interpreted as carrier generation by exciton dissociation induced by the electric field, resulting in the partial melting of the excitonic phase and metallization. The presence of a distinct threshold electric field for carrier generation indicates exciton dissociation via quantum-tunnelling processes. The spectral change due to metallization by the electric field is significantly different from that due to the strong optical excitation across the gap, which can be explained by the different melting mechanisms of the excitonic phase in the two types of excitations.

# I. INTRODUCTION

Recent advances in femtosecond laser technology have enabled the generation of nearly monocyclic intense terahertz pulses with electric-field amplitudes exceeding 1 MV/cm [1-3]. The use of such pulses to control the electronic states of solids has been the subject of much research [4-9]. A typical example of a phase transition utilizing the electric field of a terahertz pulse is the insulator–metal transition [10-12]. The first report was on $VO_2$, which is a Peierls insulator [10]. In this compound, the localized carriers present in the original compound are released and accelerated by a terahertz electric field. The resulting heating effect causes a transition to a metal. Researchers have also reported that in Mott insulators of organic molecular compounds, $\kappa$-$(ET)_2Cu[N(CN)_2]Br$ (ET: bis(ethylenedithio)tetrathiafulvalene) [11] and ET-$F_2TCNQ$ ($F_2TCNQ$: difluorotetracyano-quinodimethane) [12], the bands are spatially tilted and electrons are transferred from the lower Hubbard band to the upper one through quantum tunneling upon irradiating a terahertz electric-field pulse, resulting in metallization. An insulator–metal transition involving this mechanism has also been observed in $VO_2$ using a multicycle mid-infrared pulse for excitation [13]. In addition, a transition from the paraelectric Mott-insulator phase to the ferroelectric charge-ordered phase has been reported for the organic molecular compound, $\kappa$-$(ET)_2Cu[N(CN)_2]Cl$, which is a typical electronic-type dielectric [14]. This compound has a two-dimensional structure consisting of molecular (ET) dimers with a single hole. The terahertz pulse causes the hole in each dimer to move simultaneously along the electric field, resulting in the formation of molecules with two different charges in each dimer and a ferroelectric charge-order state. Thus, although several successful examples of electronic phase transitions induced by terahertz electric-field pulses have been reported, they are not numerous because the change in potential energy provided to an electronic system by the electric field of a terahertz pulse is not very large.

In this study, we attempted field-induced metallization of an excitonic insulator to search for a novel phase transition using a terahertz electric-field pulse. As shown in Figs. 1(a) and (b), an excitonic insulator is a material in which electron–hole pairs are stabilized by Coulomb attractive interactions when they are generated in a semi-metal with a small band overlap or a semiconductor with a small band gap. When an excitonic insulator is irradiated with an intense terahertz pulse, excitons are expected to dissociate, and a metallic state appears. Although the exciton insulator concept was theoretically proposed in the 1960s [15-17], no candidate has yet been discovered. Recently, however, candidate materials for excitonic insulators have been identified in transition metal chalcogenides [18-21] and cobalt oxides [22,23]. In particular, the transition metal chalcogenides 1$T$-TiSe$_2$ [19,20] and Ta$_2$NiSe$_5$ [21] have been actively studied in recent years because their excitonic insulator phases appear simply by lowering their temperatures. The phase transition mechanism of 1T-TiSe$_2$ is complicated because a charge density wave is formed in the excitonic insulator phase. By contrast, Ta$_2$NiSe$_5$ has a direct gap and does not form a charge density wave [21]. Therefore, in this study, we focused on Ta$_2$NiSe$_5$, which undergoes a simple transition to the excitonic insulator phase.

Figures 1(c) and (d) schematically show the arrangements of Ta and Ni ions in the *ac* and *ab* planes of Ta$_2$NiSe$_5$, respectively. The material has a quasi-one-dimensional (1D) structure consisting of two Ta chains and one Ni chain along the *a*-axis [24]. The Se ions are octahedrally coordinated around the Ta ions and tetrahedrally coordinated around the Ni ions. At high temperatures, Ta$_2$NiSe$_5$ has an orthorhombic structure [Fig. 1(c)], which is metallic above 550 K and semi-metallic below 550 K [25]. Below 328 K, Ta ions are displaced to form a monoclinic structure, and the electrical conductivity becomes semiconducting [25,26]. Recent angle-resolved photoemission spectroscopy (ARPES) measurements indicated the formation of an excitonic insulator state in the low-temperature phase of Ta$_2$NiSe$_5$, as shown in Fig. 1(b) [21,27]. Band calculations indicated that the conduction band consists of doubly degenerate 5d orbitals of Ta ions and the valence band consists of hybridized 3d orbitals of Ni ions and 4p orbitals of Se ions [28]. Furthermore, the electrons in the Ta-5d orbitals and the holes in the hybridized Ni-3d and Se-4p orbitals form excitons, which condense to produce a bandgap [29]. Very recently, the electronic state change upon photoexcitation in the excitonic insulator phase of $Ta_2NiSe_5$ and $Ta_2Ni_{1-x}Co_xSe_5$ ($x = 0.10$) was investigated using time-resolved ARPES (Tr-ARPES) [30-39], pump-probe reflection and transmission spectroscopy [40-47], pump-probe Raman and photoluminescence spectroscopies [48], and theoretical calculations [49-57]. These results confirmed the photoinduced metallization of the excitonic phase [31-35,37,38,46]. The aim of this study was to investigate the possibility of metallization of the exciton insulator phase in $Ta_2NiSe_5$ by applying a strong electric field of a terahertz pulse to dissociate the electron–hole pairs of condensed excitons. For this purpose, we performed terahertz pump mid-infrared (MIR) and near-infrared (NIR) reflection probe spectroscopy on $Ta_2NiSe_5$.

In general, when a weak electric field is applied to a semiconductor or an insulator, the symmetry of the system can be broken, resulting in the hybridization of wave functions with originally opposite symmetries or odd and even parities. Therefore, by analyzing the change in the optical spectrum owing to an applied electric field, one can obtain information about the originally forbidden electronic transitions that are not observed in the linear absorption spectrum. This technique is called electro-reflectance or absorption spectroscopy and has been applied to various types of semiconductors and insulators [58-67]. More recently, electro-reflectance spectroscopy by applying the electric field of a terahertz pulse to a material has been proposed and utilized to characterize the nature of the one-photon allowed excited state detected in the steady-state optical spectrum and that of the one-photon forbidden excited state in low-dimensional Mott insulators [68-70]. In the present study, we first applied this electro-reflectance spectroscopy technique using a weak electric field from a terahertz pulse to $Ta_2NiSe_5$. This approach is effective for characterizing the optical transition inherent in the low-temperature exciton-condensed phase, which should be observed at the absorption edge in the steady-state reflectivity and absorption spectra. Next, we investigated the changes in the wide energy range of the reflectivity spectrum of $Ta_2NiSe_5$ excited by a terahertz pulse with a strong electric field. The absorption intrinsic to condensed excitons at the band edge decreased, and the Drude response appeared at lower energies. This finding demonstrates that the terahertz electric field dissociates excitons to produce carriers and metallization occurs. Furthermore, the spectral changes and their temporal characteristics reflecting this electric field-induced insulator–metal transition were found to be significantly different from those observed in previously reported photoinduced insulator–metal transitions. Based on the experimental results, this paper discusses the differences between the mechanisms of metallization induced by a terahertz electric field and photoexcitation far beyond the gap.

## II. EXPERIMENTAL DETAILS

### A. SAMPLE PREPARATIONS

Single crystals of $Ta_2NiSe_5$ were prepared using the chemical vapor transport method [24]. The broad plane is the *ac* plane on which all optical measurements were performed. The typical dimensions of this plane are 1.2 mm in the *a*-direction and 0.5 mm in the *c*-direction. The typical crystal thickness was 20 µm.

### B. POLARIZED REFLECTION SPECTROSCOPY IN STEADY STATE

The polarized optical reflectivity (*R*) spectra of single crystals of $Ta_2NiSe_5$ were measured using a specially designed spectrometer with a 25-cm-grating monochromator at 0.5–5.0 eV and a Fourier-transform infrared spectrometer at 0.012–1.2 eV. Each spectrometer was equipped with an optical microscope. The temperature of a single crystal was varied by attaching it to the sample holder of a conducting cryostat with a diamond window.

## C. TERAHERTZ PUMP MID-INFRARED AND NEAR-INFRARED REFLECTIVITY PROBE SPECTROSCOPY

In the terahertz pulse pump optical reflectivity probe experiments, two systems I and II were used, as illustrated in Figs. 2(a) and (b). To perform electro-reflectance spectroscopy using terahertz electric fields, low-frequency terahertz pulses obtained in system I are appropriate. In this system, the light source was a Ti:Al$_2$O$_3$ regenerative amplifier (RA) with a temporal width of 90 fs, photon energy of 1.55 eV, repetition rate of 1 kHz, and pulse energy of 5 mJ. The RA output was divided into two beams. One was used for the generation of terahertz pump pulses through optical rectification in a LiNbO$_3$ crystal using the pulse-front-tilting method, as illustrated in Fig. 2(a) [1,2]. The magnitude of the terahertz electric field was changed using two wire-grid polarizers inserted into the optical path of the terahertz pulses. The second was introduced into an optical parametric amplifier (OPA) to obtain probe pulses with photon energies ranging from 0.12 to 0.47 eV.

A terahertz pulse generated by LiNbO$_3$ was not sufficiently strong to produce exciton dissociation in the excitonic phase, and one generated by an organic nonlinear optical crystal of 4-N,N-dimethylamino-4'-N'-methyl-stilbazolium 2,4,6-trimethylbenzenesulfonate (DSTMS) in system II was used, as illustrated in Fig. 2(b) [3,12,71]. In this system, the light source was another Ti:Al$_2$O$_3$ RA with a temporal width of 100 fs, photon energy of 1.55 eV, repetition rate of 1 kHz, and pulse energy of 7 mJ. The output is divided into two beams. One is introduced into the OPA to obtain an NIR pulse (0.8 eV) that is employed to excite the DSTMS crystal. A terahertz pulse with a high electric field amplitude is emitted by the DSTMS crystal. The magnitude of the terahertz electric field was changed by adjusting the intensity of the incident 0.8-eV pulse with a neutral density filter. The other beam from the RA output is introduced into another OPA to obtain probe pulses with photon energies ranging from 0.09 to 0.78 eV.

The temporal waveform of a terahertz electric field pulse, $E_{\text{THz}}(t)$, was measured by electro-optical sampling with a 0.2-mm-thick (110)-oriented ZnTe crystal and a 0.1-mm-thick (110)-oriented GaP crystal for systems I and II using LiNiO$_3$ and DSTMS crystals, respectively [8,72]. Just before introducing a terahertz pulse into the nonlinear optical crystal, its intensity was adjusted using Si plates. The time origin for the terahertz pump optical probe experiments was set at the maximum of the terahertz electric field amplitude. The time difference between the pump and probe pulses, $t$, was controlled by varying the path length of the probe pulse. The polarization of all the pulses was parallel to the 1D stacking axis $a$.

To perform low-temperature measurements, a single-crystal sample was mounted and cooled in a cryostat with optical windows made of diamond with a thickness of 400 μm. The maximum electric-field amplitude of the terahertz pulses in the cryostat was approximately 300 kV/cm for the system using LiNbO$_3$ and approximately 1.9 MV/cm for the system using DSTMS.

## III. RESULTS

### A. STEADY-STATE OPTICAL SPECTRA

Figure 1(e) shows the reflectivity ($R$) spectra with the electric field $E$ parallel to the $a$-axis ($E//a$) in Ta$_2$NiSe$_5$. At 340 K above $T_c$ and 300 K just below $T_c$, $R$ was high at low energies below 0.1 eV, indicating that the material was in the high conductivity state. On the other hand, at 10 K, $R$ decreases with decreasing photon energy, reflecting the insulator state. Figure 1(f) represents the spectra of the real part of the complex conductivity, $\sigma_1$, that is, of the optical conductivity, which were obtained by Kramers-Kronig (KK) transformation of the $R$ spectra. To perform the KK transformation, the $R$ spectrum was extrapolated in the energy regions below 0.012 eV and above 5.0 eV in which $R$ was not measured. Above 5.0 eV, $R$ was assumed to be constant from 5.0 eV to 13 eV, and extrapolated using the relation $R \propto \omega^{-4}$ above 13 eV.

Below 0.012 eV, $R$ was extrapolated assuming the Hagen–Rubens relation at 340 K and 300 K since the material was conductive, while $R$ was assumed to be constant and set to the value at 0.012 eV at 10 K. In the $\sigma_1$ spectra at 340 K and 300 K, a high spectral intensity occurred at low energies below 0.1 eV. On the other hand, at 10 K, the $\sigma_1$ spectrum shows a clear optical gap. These spectral features of $R$ and $\sigma_1$ are almost in agreement with those previously reported [26,46,73-75]. The blue line in Fig. 3(a) shows the magnified $\sigma_1$ spectrum at 10 K, which has a peak around 0.4 eV and a shoulder-like structure just below the peak. The second-order energy derivative of the $\sigma_1$ spectrum is depicted in Fig. 3(b). The dip energies of 0.405 eV and 0.36 eV in this spectrum correspond to those two structures. As seen in the $\sigma_1$ spectra in Fig. 1(e), the peak structure around 0.4 eV is observed at three different temperatures, whereas the shoulder structure at 0.36 eV is evident only at 10 K. Therefore, the latter structure can be attributed to the transition inherent to the exciton-condensed state. The origin of this structure is discussed in more detail in Subsection III-B.

## B. TERAHERTZ PUMP OPTICAL REFLECTIVITY PROBE SPECTROSCOPY: THIRD-ORDER NONLINEAR OPTICAL RESPONSE

To obtain information about the origins of the structures appearing in the $\sigma_1$ spectrum, we applied terahertz pump reflectivity probe spectroscopy [Fig. 2(a)] to Ta$_2$NiSe$_5$ at 10 K and investigated the change in the $R$ spectrum owing to the electric field. In this measurement, terahertz pump pulses were generated by a second-order nonlinear optical crystal, LiNbO$_3$ [Fig. 2(a)] [1,2]. Figure 2(b) shows the electric field waveform of the terahertz pulse, which exhibits a nearly monocyclic electric field. The maximum electric field, $E_{\text{THz}}(0)$, is approximately 0.3 MV/cm. The blue line in Fig. 2(c) shows the Fourier power spectrum of the electric-field waveform, which has a peak at 0.75 THz. Figure 2(d) depicts the $\sigma_1$ spectrum along the $a$-axis in the terahertz region of Ta$_2$NiSe$_5$, which was reproduced using the parameters of the Lorentz oscillators deduced from the fitting analysis of the previously reported experimental reflectivity spectrum [76]. Each peak was attributed to an infrared-active phonon. One phonon mode exists with a peak at 1.14 THz in the spectral range of the terahertz pulse, but its absorption intensity is so weak that $Ta_2NiSe_5$ can be considered transparent to this pulse.

Figures 4(a), (b), and (c) (green circles) present the time characteristics of the terahertz electric field-induced reflectivity changes, $\Delta R(t)$, at 0.17 eV, 0.36 eV, and 0.45 eV, respectively. Positive $\Delta R(t)$ signals appear at 0.17 eV and 0.45 eV, whereas a negative $\Delta R(t)$ signal occurs at 0.36 eV. The time dependences of all the signals are almost in agreement with the square of the electric field waveform $E_{\mathrm{THz}}(t)$ [Fig. 2(c)], $[E_{\mathrm{THz}}(t)]^2$, as indicated by the blue lines in the same images, demonstrating that the $\Delta R(t)$ signals are proportional to $[E_{\mathrm{THz}}(t)]^2$. The time characteristics of $\Delta R(t)$ at 0.36 eV for various electric-field amplitudes, $E_{\mathrm{THz}}(0)$, are presented in Fig. 4(d). The dependence of the change in reflectivity at the time of origin, $\Delta R(0\ \mathrm{ps})$, on $E_{\mathrm{THz}}(0)$ is shown in Fig. 4(e). The magnitudes of $|\Delta R(0\ \mathrm{ps})|$ are again proportional to $[E_{\mathrm{THz}}(0)]^2$ up to $E_{\mathrm{THz}}(0) \sim 0.3$ MV/cm, as shown by the red line. These results demonstrate that the observed reflectivity changes are due to a third-order nonlinear optical response expressed by the following equation:

$$\Delta P(\omega) = 3\varepsilon_0 \chi^{(3)}(-\omega;0,0,\omega)[E_{\mathrm{THz}}(0)]^2 E(\omega). \tag{1}$$

Here, $\Delta P(\omega)$ is the third-order nonlinear polarization due to the terahertz electric field $E_{\mathrm{THz}}(\omega\sim0)$ and the electric field of the probe pulse, $E(\omega)$. $\chi^{(3)}(-\omega;0,0,\omega)$ is the third-order nonlinear susceptibility that determines the magnitude of $\Delta P(\omega)$.

To obtain more detailed information about the mechanisms of these reflectivity changes, the time characteristics of $\Delta R(t)$ were measured at various probe energies. The obtained $\Delta R(0\ \mathrm{ps})$ spectrum is plotted as red circles in Fig. 5(b). As the probe energy increases from 0.1 eV, positive, negative, and positive reflectivity changes appear around the peak in the original reflectivity spectrum depicted in Fig. 5(a). To investigate this spectral change in more detail, the changes of the complex dielectric constant, $\Delta\tilde{\varepsilon}$, were calculated from the $\Delta R(0\text{ ps})$ spectrum by the KK transform. The $\Delta R(0\text{ ps})$ spectrum outside the measurement region was extrapolated as shown by the red dashed line in Fig. 5(b): for the low (high) energy side, we extrapolated a straight line from the lower (higher) energy bound of the measurement to $\Delta R(0\text{ ps}) = 0$ at 0.1 eV (0.5 eV). KK transformation was performed on the $R + \Delta R$ spectrum. The red lines in Figs. 5(d) and (e) represent the spectra of the real part, $\Delta\varepsilon_1$, and imaginary part, $\Delta\varepsilon_2$, of $\Delta\tilde{\varepsilon}$ obtained from this analysis, respectively. In the $\Delta\varepsilon_2$ spectrum, which corresponds to the absorption change, we can see positive, negative, and positive structures resembling the $\Delta R$ spectrum.

As a mechanism for the change of an $\varepsilon_2$ spectrum by an external electric field in a semiconductor, the Franz-Keldysh effect is well known [77,78]. This phenomenon originates from the electron wavefunction penetrating the forbidden gap due to the electric field. Consequently, the absorption edge shifts to the low-energy side, and an oscillatory structure that depends on the electric field strength appears on the high-energy side of the absorption edge. To check this possibility, we calculated the $\Delta\tilde{\varepsilon}$ spectrum due to the Franz-Keldysh effect. However, the experimental $\Delta\tilde{\varepsilon}$ spectrum could not be reproduced, as reported in Appendix A.

Based upon the electric-field dependence discussed above, it is reasonable to consider that the $\Delta\tilde{\varepsilon}$ spectrum is due to the third-order nonlinear optical effect. The positive–negative–positive structure observed in the $\Delta\varepsilon_2$ spectrum in Fig. 5(e) is similar to those obtained from the electro-reflectance spectroscopy of 1D Mott insulators, such as halogen-bridged nickel-chain compounds, [Ni(chxn)$_2$X]X$_2$ (chxn = cyclohexanediamine, and $X$ = Cl, Br), cuprates $A_2$CuO$_3$ ($A$ = Sr and Ca), and an organic molecular compound, bis(ethylenedithio)-tetrathiafulvalene-difluorotetracyanoquinodimethane (ET-F$_2$TCNQ) [66-68,70]. In these 1D Mott insulators, the analyses of the $\Delta\varepsilon_2$ spectra revealed that two excited states exist at the absorption edge: the lowest one-photon allowed excited state $|1\rangle$ with odd parity and the one-photon forbidden excited state |2⟩ with even parity located just above |1⟩. When an electric field is applied, these two states hybridize. At that time, state |1⟩ is shifted to a lower energy, and the positive and negative absorption changes in $\Delta\varepsilon_2$ emerge from low to high energy. On the other hand, state |2⟩ gains oscillator strength, and an induced absorption, i.e., a positive absorption change of $\Delta\varepsilon_2$, appears on the high-energy side. This phenomenon results in positive, negative, and positive changes of $\Delta\varepsilon_2$ with increasing energy. Indeed, the $\Delta\varepsilon_2$ spectra in the 1D Mott insulators could be reproduced by a three-level model that includes the ground state |0⟩, one-photon allowed excited state |1⟩, and one-photon forbidden excited state |2⟩. Therefore, we analyzed the $\Delta\tilde{\varepsilon}$ spectrum in Ta$_2$NiSe$_5$ shown in Figs. 5(d) and (e) using a similar three-level model.

The electric-field effect was expected to be significant in transitions specific to the exciton-condensed state, as indicated by the orange arrow in Fig. 1(b). As seen in Fig. 5(d), the center of gravity of the positive, negative, and positive changes of the $\Delta\varepsilon_2$ spectrum is located near the shoulder structure around 0.36 eV on the low-energy side of the main peak in the $\sigma_1$ spectrum in Fig. 5(c). Therefore, this shoulder structure can be reasonably attributed to the one-photon-allowed excited state |1⟩ with the odd parity inherent to the exciton-condensed state. With these considerations in mind, the shoulder structure at 0.36 eV in the $\sigma_1$ spectrum was analyzed using the following Lorentzian function:

$$\sigma_1(\omega) = \omega\varepsilon_0\varepsilon_2(\omega) = \omega\varepsilon_0 \frac{Ne^2}{\hbar} \langle 0|x|1\rangle^2 \text{Im}\left(\frac{1}{\omega_1 - \omega - i\gamma_1} + \frac{1}{\omega_1 + \omega + i\gamma_1}\right). \quad (2)$$

Here, $\hbar\omega_1$ and $\hbar\gamma_1$ are the energy and width of the excited state |1⟩ and $\langle 0|x|1\rangle$ is the transition dipole moment between the ground state |0⟩ and the excited state |1⟩. In the analysis, we set $\hbar\omega_1$ to 0.36 eV, as estimated from the second energy derivative spectrum shown in Fig. 3(b). As a result of the fitting procedure, the structure of $\sigma_1$ around the absorption edge was reproduced, as indicated by the orange lines in Figs. 3(b) and 5(c). This component can be

ascribed to the transition of the exciton state [the orange arrow in Fig. 1(b)] and is labelled as $\sigma_{1\text{ex}}(\omega, \omega_1)$. $\omega_1$ in $\sigma_{1\text{ex}}(\omega, \omega_1)$ denotes the frequency of this transition. The parameters used are $\langle 0|x|1 \rangle = 1.75$ Å and $\hbar\gamma_1 = 0.065$ eV, which are listed in Table I. The brown line in Fig. 5(c) shows the spectrum obtained by subtracting the $\sigma_{1\text{ex}}$ from the overall $\sigma_1$ spectrum represented by the blue line. This structure following the high-energy side is inferred to be due to the interband transition indicated by the brown arrow in Fig. 1(b) and labelled as $\sigma_{1\text{b}}$. In a previous study [73], the presence of two transitions was also considered in the $\sigma_1$ spectrum, and the structures were interpreted as corresponding to excitons with principal quantum numbers $n = 1, 2$ or as arising from the lifting of the degeneracy of the conduction band consisting of Ta 5d orbitals. Although we also considered two excited states that are characteristic of the exciton-condensed state in this study, our interpretation differs from that of previous studies.

Next, we will explain the analysis of the $\Delta\tilde{\varepsilon}$ spectrum. From Eq. (1), the following relationship between $\Delta\tilde{\varepsilon}$ and $\chi^{(3)}(-\omega; 0,0, \omega)$ can be derived:

$$\Delta\tilde{\varepsilon}(\omega) = 3\chi^{(3)}(-\omega; 0,0, \omega)[E_{\text{THz}}(0)]^2. \tag{3}$$

This relationship shows that $\Delta\varepsilon_1$ and $\Delta\varepsilon_2$ correspond to $\text{Re}\chi^{(3)}$ and $\text{Im}\chi^{(3)}$, respectively. The right vertical axes of Figs. 5(d) and (e) show the values of $\text{Re}\chi^{(3)}$ and $\text{Im}\chi^{(3)}$ calculated using Eq. (3). In this calculation, the local values of the terahertz electric field inside the crystal were used by considering the Fresnel reflection and dielectric constant (Appendix B). In the three-level model, the main term of the expression for $\chi^{(3)}(-\omega; 0,0, \omega)$, denoted as $\chi^{(3)}(-\omega; 0,0, \omega)_{\text{main}}$, is given by the following equation [66-69]:

$$\chi^{(3)}(-\omega; 0,0, \omega)_{\text{main}} = \frac{Ne^4}{3\varepsilon_0\hbar^3} \langle 0|x|1 \rangle \langle 1|x|2 \rangle \langle 2|x|1 \rangle \langle 1|x|0 \rangle \left[\frac{1}{(\omega_1 - \omega - i\gamma_1)^2(\omega_2 - \omega - i\gamma_2)}\right] \tag{4}$$

Here, $\hbar\omega_2$ and $\hbar\gamma_2$ are the energy and width of the one-photon forbidden state $|2\rangle$ with even parity, and $\langle 1|x|2\rangle$ is the transition dipole moment between $|1\rangle$ and $|2\rangle$. For the parameters of $\omega_1$, $\gamma_2$, and $\langle 0|x|1\rangle$, we can use the values obtained from the analysis of the $\sigma_1$ spectrum, so the fitting parameters are $\langle 1|x|2\rangle$, $\hbar\omega_2$, and $\hbar\gamma_2$. In the actual fitting analysis, we used the complete expression of $\chi^{(3)}(-\omega; 0,0,\omega)$ consisting of 12 terms to fit both the Re$\chi^{(3)}$ and Im$\chi^{(3)}$ spectra simultaneously (Appendix B). The green lines in Figs. 5(d) and (e) represent the fitting curves, which reproduce the experimental data shown by the red line. The fitting parameters used are $\hbar\omega_2 = 0.396$ eV, $\hbar\gamma_2 = 0.097$ eV, and $\langle 1|x|2\rangle = 0.95$ Å, which are also listed in Table I. To compare the properties of the even-parity state $|2\rangle$ with those of the odd-parity state $|1\rangle$, the spectrum of $|2\rangle$ is shown by the gray broken line on a normalized scale in Fig. 5(c).

The energy difference between excited states $|1\rangle$ and $|2\rangle$ in the three-level model is only 36 meV; that is, their energies are very close. In addition, the energy of the latter is still lower than that of the interband transition $\sigma_{1b}$ [brown line in Fig. 5(c)]. Therefore, states $|1\rangle$ and $|2\rangle$ can reasonably be assumed to be specific to the exciton-condensed phase as well. Considering that the exciton-condensed state in Ta$_2$NiSe$_5$ is generated by the stabilization of electrons in Ta and holes in Ni by Coulomb attractive interactions, the lowest excited state is ascribed to the transition in which electrons in Ta move to Ni or, equivalently, holes in Ni move to Ta. This transition is schematically illustrated in the hole picture shown in Fig. 5(f). Here, we simply consider a 1D system in which Ta and Ni ions are arranged alternately; the 1D system was originally a narrow-gap semiconductor, as shown in the upper panel of Fig. 5(f). The blue and red arrows represent the spins of the holes, and the on-site Coulomb repulsion energy at each site is omitted in this figure. The lower panel of Fig. 5(f) shows the excitonic insulator phase, in which a hole in each Ta site is transferred to a Ni site and stabilized by the Coulomb interaction $V_{\text{ex}}$, as indicated by the green arrow. The lowest electronic excitation in this phase was the transfer of a hole from a Ni site to a neighboring Ta site. Assuming that the excited state in which a hole in a Ni site moves to the right (left) is denoted as $|\rightarrow\rangle$ ($|\leftarrow\rangle$), the excited states consist of linear combinations of the two excited states, an antisymmetric state with odd parity, $\frac{1}{\sqrt{2}}(|\rightarrow\rangle - |\leftarrow\rangle)$, and a symmetric state with even parity, $\frac{1}{\sqrt{2}}(|\rightarrow\rangle + |\leftarrow\rangle)$, as shown in Fig. 5(g). Electrons and holes exist at different sites in the exciton-condensed state. Therefore, these two excited states can be expected to be close to each other. This situation resembles the case of odd- and even-parity exciton states near the band edge in half-filled 1D Mott insulators, where they are almost degenerate in energy.

## C. TERAHERTZ PUMP OPTICAL REFLECTIVITY PROBE SPECTROSCOPY: ELECTRIC-FIELD DEPENDENCE OF REFLECTIVITY CHANGES

The main objective of this study was to irradiate the excitonic insulator phase of $Ta_2NiSe_5$ with a strong electric field pulse to dissociate excitons, melt the excitonic insulator phase, and induce metallization while elucidating the dynamics of these processes. As mentioned in Subsection III-B, when an electric-field pulse with an amplitude of up to 0.3 MV/cm is irradiated onto $Ta_2NiSe_5$, third-order nonlinear optical responses occur, but the exciton-condensed state does not melt. To apply the higher electric-field amplitude of a terahertz electric-field pulse, we used the pulse generated by exciting an organic nonlinear optical crystal of DSTMS with an NIR pulse. The typical electric field waveform and spectrum of the obtained terahertz pulse are indicated by the red lines in Figs. 2(d) and (e), respectively. The central frequency of the pulse was approximately 3 THz, and the bandwidth extends from 1.5 to 4.5 THz. Therefore, the IR-active vibrational modes around 3.4 THz, 4.1 THz, and 4.5 THz observed in the $\sigma_1$ spectrum shown in Fig. 2(f) may be excited. Corresponding to the broad spectral width, the electric field of the pulse is localized in a narrower time domain of approximately 1 ps compared to that emitted by LiNbO$_3$ [blue lines in Fig. 2(c)]. The maximum electric-field amplitude that can be applied to a crystal in a cryostat is approximately 1.9 MV/cm.

Figures 6(b) and (c) exhibit, respectively, the time characteristics of terahertz electric field-induced reflectivity changes, $\Delta R(t)$, for various electric-field amplitudes, $E_{\text{THz}}(0)$, at 0.39 eV, at which the reflectivity spectrum shows a peak structure, and at 0.11 eV, at which the increase in reflectivity upon metallization is predicted. In Fig. 6(a), the square of the electric field waveform $E_{\text{THz}}(t)$, $[E_{\text{THz}}(t)]^2$ is presented. Because Ta$_2$NiSe$_5$ has inversion symmetry, the response can be expected to be proportional to $[E_{\text{THz}}(t)]^2$, as in the weak electric field case discussed in Subsection III-B. For $E_{\text{THz}} = 0.49$ MV/cm, a sharp decrease in reflectivity was observed near the time origin owing to the aforementioned third-order nonlinear optical response.

The electric field of the terahertz pulse generated by the DSTMS oscillates rapidly, as shown in Fig. 2(e), and its square, $[E_{\text{THz}}(t)]^2$, oscillates even more rapidly, as depicted in Fig. 6(a). Therefore, the rapid response that follows the waveform of $[E_{\text{THz}}(t)]^2$ cannot be captured in real time using a probe pulse with a temporal width of ∼100 fs, as employed in this study. As expected, however, the reflectivity change induced by the terahertz pulse with an amplitude of 0.49 MV/cm indeed almost matches the envelope of $[E_{\text{THz}}(t)]^2$ [Figs. 6(a) and (b)]. The red circles in Fig. 6(d) show the dependence of the reflectivity changes at the time origin, $\Delta R(0 \text{ ps})$, on the terahertz electric-field amplitude, $E_{\text{THz}}(0)$. The magnitude of $-\Delta R(0 \text{ ps})$ is proportional to $[E_{\text{THz}}(t)]^2$ for electric-field amplitudes up to approximately 0.7 MV/cm, confirming that the response around the time origin is due to the third-order nonlinear optical responses. Upon further increasing $E_{\text{THz}}(0)$, a significant change in the time characteristics of $\Delta R(t)$ also occurs. As shown in Fig. 6(b), when $E_{\text{THz}}(0)$ exceeds 1 MV/cm, another decrease in reflectivity appears before the reflectivity changes near the time origin owing to the third-order nonlinear optical response recovers. This decrease in reflectivity increases nonlinearly with increasing $E_{\text{THz}}(0)$. At $E_{\text{THz}}(0)$ of 1.9 MV/cm, the decrease in reflectivity reaches a maximum near 0.7 ps and then slowly recovers. This response suggests that another large electronic state change occurs because of the terahertz electric field, rather than the third-order nonlinear optical response. As shown in Fig. 5(c), the reflectivity change at 0.39 eV corresponds mainly to the change in the transition to the excited state $|1\rangle$, so that the decrease in reflectivity at this energy will reflect a decrease in the number of condensed excitons.

To obtain information about the origin of this response, the dependence of the reflectivity decrease $-\Delta R(t)$ at 1 ps, $-\Delta R(1 \text{ ps})$, at 0.39 eV on $E_{\text{THz}}(0)$ is plotted in Fig. 6(e). $-\Delta R(1 \text{ ps})$ shows a clear threshold behavior, indicating a process in which electrons and holes in the exciton-condensed state dissociate due to the terahertz electric field. In fact, at electric fields above 0.8 MV/cm where $-\Delta R(0)$ [red circles in Fig. 6(d)] tends to saturate, the third-order nonlinear optical response can be inferred to be suppressed by the exciton dissociations. This electric field-induced exciton dissociation can be considered as a process in which electrons that constitute excitons in the upper part of the valence band tunnel to the lower part of the conduction band or holes that constitute excitons in the lower part of the conduction band tunnel to the upper part of the valence band. The process in which the band is spatially tilted by an intense electric field and electrons (holes) in the valence band (conduction band) tunnel to the conduction band (valence band) has been studied in detail in 1D Mott insulators [12,79].

Theoretical studies on 1D Mott insulators have shown that the probability of carrier generation via the tunneling process, $\Gamma$, is proportional to $E_{\text{THz}} \exp(-\pi E_{\text{th}}/E_{\text{THz}})$ [79], where $E_{\text{THz}}$ is the amplitude of the external electric field and $E_{\text{th}}$ is the threshold electric field. A recent experimental study on a typical 1D Mott insulator, ET-F$_2$TCNQ, demonstrated that electric field-induced carrier generation occurs according to this relationship. The 1D exciton-condensed state is produced by strong electron interactions; thus, a response to an electric field similar to that observed in 1D Mott insulators is possible, as schematically illustrated in Fig.

7(a). Therefore, fitting analysis was performed on the electric field dependence of $-\Delta R$ at 1 ps and at 0.39 eV [red circles in Fig. 6(e)] using $E_{\text{THz}}\exp(-\pi E_{\text{th}}/E_{\text{THz}})$. When $E_{\text{th}}$ is set to 0.8 MV/cm, the fitting curve reproduces the experimental results well, as shown by the green line in Fig. 6(e), indicating that the exciton dissociations indeed occur via quantum tunneling processes.

As seen in Fig. 6(b), the large decrease in reflectivity at 0.39 eV due to carrier generation seems to start at approximately 0.1 ps and reaches a maximum at approximately 0.7 ps. To explain the origin of this delayed response, interpretation of the previously reported response to the 1.55-eV optical excitation far beyond the energy gap provides important information. In Ta$_2$NiSe$_5$, a metallic response was observed when the exciton-condensed state was strongly excited by a 1.55 eV pulse. The results revealed a finite time delay in the manifestation of the metallic response, which was interpreted as the time required for the generated carriers to reach the plasma state [46,80]. In the case of carrier generation by the terahertz electric field, a finite time is also probably required to reach a metallic plasma state, which will be responsible for the time delay in the $-\Delta R(t)$ signals.

The response of the carriers generated by the electric field is expected to be prominent in the MIR region. Therefore, the time evolution of $\Delta R(t)$ at 0.11 eV, which is below the optical gap, was measured at various electric field amplitudes $E_{\text{THz}}(0)$, as shown in Fig. 6(c). At this photon energy, the time resolution is slightly worse than that at 0.39 eV because the temporal width of the probe pulse is extended to approximately 200 fs [12]. When $E_{\text{THz}}(0)$ is 0.33 MV/cm, negative $\Delta R(t)$ appears near the time origin, but the response disappears as the terahertz electric field decays. However, when $E_{\text{THz}}(0)$ is increased to 0.8 MV/cm or higher, a negative $\Delta R(t)$ appears in the time domain after the terahertz field decays. Furthermore, as $E_{\text{THz}}(0)$ exceeds 1.4 MV/cm, a positive $\Delta R(t)$ appears near 1 ps after excitation. This positive $\Delta R(t)$ quickly decreases and turns negative around 1.5 ps, and the negative $\Delta R(t)$ decays and approaches zero after 2 ps. For $E_{\text{THz}}(0)$ of 1.4 MV/cm or higher, $\Delta R(t)$ becomes positive again after 4 ps.

To obtain information about the origin of these responses, the $E_{\text{THz}}(0)$ dependence of $\Delta R(t)$ at the time origin and 1 ps after excitation, $\Delta R(0\ \text{ps})$ and $\Delta R(1\ \text{ps})$, are indicated by blue triangles in Figs. 6(d) and (f), respectively. The $-\Delta R(0\ \text{ps})$ signal is proportional to $[E_{\text{THz}}(0)]^2$ in the low electric field region. In contrast, when $E_{\text{THz}}(0)$ increases above 0.8 MV/cm, it deviates from the proportional relationship with $[E_{\text{THz}}(0)]^2$ and tends to saturate. This behavior resembles the $E_{\text{THz}}(0)$ dependence of $\Delta R(0\ \text{ps})$ at 0.39 eV [red circles in Fig. 6(d)]. These results indicate that the $-\Delta R(0\ \text{ps})$ signal is also due to a third-order nonlinear optical response.

On the other hand, $\Delta R(1\ \text{ps})$ at 0.11 eV, where the terahertz electric field shown in Fig. 6(f) completely disappears, exhibits a complex behavior with respect to $E_{\text{THz}}(0)$. Accordingly, almost no signal exists when $E_{\text{THz}}(0)$ is below 0.5 MV/cm, which corresponds well to the behavior of the electric-field dependence of $\Delta R(1\ \text{ps})$ at 0.39 eV shown in Fig. 6(e). However, when $E_{\text{THz}}(0)$ exceeds 0.5 MV/cm, a negative reflectivity change appears. Furthermore, as $E_{\text{THz}}(0)$ increases further, $\Delta R(1\ \text{ps})$ increases rapidly at 1.2 MV/cm and the sign of the signal changes from negative to positive at 1.5 MV/cm. This positive $\Delta R(1\ \text{ps})$ signal can be interpreted as the appearance of spectral intensity due to the dissociation of excitons and the generation of electron and hole carriers by the electric field. However, the reflectivity change (1 ps) at 0.11 eV in the intermediate electric-field region from 0.5 MV/cm to 1.4 MV/cm in Fig. 6(f) is different from that at 0.39 eV [Fig. 6(e)]. To clarify this difference, the temporal evolution of the spectra must be examined, as discussed in Subsection III-D.

## D. TERAHERTZ PUMP OPTICAL REFLECTIVITY PROBE SPECTROSCOPY: TIME DEPENDENCE OF SPECTRAL CHANGES

To clarify the origin of the complex electric-field dependence of the reflectivity changes $\Delta R(t)$ at 0.11 eV as seen in Figs. 6(c) and (f), the time characteristic of $\Delta R(t)$ induced by a terahertz pulse with an electric field amplitude $E_{\text{THz}}(0)$ of 1.65 MV/cm was investigated for a probe energy range from 0.09 eV to 0.8 eV. $\Delta R(t)$ at several representative probe energies is shown in Fig. 8(a). The measurement temperature was 10 K, like in the measurements mentioned above. Using these results, we constructed the $\Delta R(t)$ spectra at $t = 0$ ps, 1 ps, and 7 ps, as shown in Fig. 8(c). Figure 8(b) presents the steady-state reflectivity spectra at 60 K as well as 10 K for which the transient reflectivity changes were measured. In the spectrum of $\Delta R(t)$ at $t = 0$ ps, $\Delta R(0 \text{ ps})$, represented by open triangles in Fig. 8(c), negative reflectivity changes appear above 0.3 eV and are centered at 0.4 eV, at which the steady-state reflectivity spectrum peaks. On the lower energy side, $\Delta R(0 \text{ ps})$ becomes positive below 0.3 eV and negative again below 0.15 eV. This $\Delta R(0 \text{ ps})$ spectrum resembles but is not exactly identical to the $\Delta R(0 \text{ ps})$ spectrum originating from the third-order nonlinear optical response obtained when the electric field amplitude is as small as 0.3 MV/cm. This difference is presumably due to carrier generation by quantum tunneling processes, in addition to the third-order nonlinear optical response. Indeed, at this electric field amplitude (1.65 MV/cm), $\Delta R(0 \text{ ps})$ is saturated at both 0.39 eV and 0.11 eV and does not follow the square of the electric field [Fig. 6(d)].

The open circles in Fig. 8(c) show the spectrum of $\Delta R(t)$ at 1 ps, $\Delta R(1 \text{ ps})$. In this spectrum, a negative signal of up to 2.5% appears around 0.4 eV, where the steady-state reflectivity spectrum peaks, and a slight positive signal is seen at 0.25 eV on the low energy side. This spectral change resembles those for the weak and strong electric fields at $t = 0$ ps shown in Figs. 5(b) and 8(c), respectively. However, as the terahertz electric field completely disappears at $t = 1$ ps, this spectral change is not due to a third-order nonlinear optical response. Below 0.15 eV, a Drude-like response appears, where $\Delta R(1 \text{ ps})$ increases monotonically toward the lower energy side. This characteristic suggests that free carriers were generated by the electric field, which was observed as a Drude response after a certain amount of time had elapsed. In this case, the decrease in reflectivity at 0.4 eV can be understood as breaching due to the carrier generation. The origin of the peak at 0.25 eV is attributed to a low-energy shift of the original transition of the excitonic phase. If carriers are generated, the Coulomb attraction that forms the exciton-condensed state is screened, which may result in a decrease in the original transition of the condensed excitons. A detailed analysis of this spectrum is presented in Section IV.

The squares in Fig. 8(c) show the spectrum of $\Delta R(t)$ at 7 ps, $\Delta R(7 \text{ ps})$, in which a negative signal appears above 0.3 eV and a positive signal appears below 0.3 eV. The time characteristics of $\Delta R(t)$ in Fig. 7(a) show that the $\Delta R$ signals observed in this time region have long lifetimes of more than 10 ps. A possible explanation for this characteristic is the effect of temperature increase in the system. To ascertain this possibility, the differential reflectivity spectrum between 60 K and 10 K, $(R_{60 \text{ K}} - R_{10 \text{ K}})$, is shown in Fig. 8(c) as an orange line, which roughly reproduces the $\Delta R$ spectrum. This indicates that the response in this time region originated from the temperature increase. In fact, the maximum temperature rise of the system was estimated to be approximately 40 K using the specific heat of $Ta_2NiSe_5$ and the energy of the terahertz pulse irradiated on the crystal. This finding is consistent with the interpretation above (see Appendix C).

Next, we discuss in more detail the time dependence of the $\Delta R(t)$ spectrum below 0.2 eV, which is shown in Figs. 8(d) and (e). At 0 ps, $\Delta R$ changes from positive to negative around 0.15 eV with decreasing energy. At 1 ps, $\Delta R$ goes from positive to slightly negative with decreasing energy, turns positive around 0.13 eV, and then monotonically increases. From 1 ps to 1.6 ps, this positive $\Delta R(1 \text{ ps})$ signal decreases with time, and the energy position at which $\Delta R(1 \text{ ps})$ changes from negative to positive shifts to lower energies. Considering that the positive $\Delta R(1 \text{ ps})$ signals below 0.13 eV are caused by free electrons and holes dissociated from excitons by the electric field-induced quantum tunneling processes, their shift to lower energy with time is caused by the decrease in carrier number.

As shown in Figs. 6(c) and 8(d), a clear time delay occurs before the Drude-like response appears. Theoretical studies indicated that the build-up time of screening by carriers generated by photoirradiation, $\tau_r$, is proportional to the inverse of the plasma frequency, $\omega_p$, which is proportional to the root of the carrier number, $N_c$. Assuming that $N_c$ is proportional to the photocarrier density, $x_{ph}$, we obtain the relation $\tau_r \propto (x_{ph})^{-\frac{1}{2}}$. In fact, the time $\tau_r$ required for the melting of the exciton insulator phase of 1T-TiSe$_2$ has been reported to decrease with increasing $x_{ph}$ and to depend on $(x_{ph})^{-\frac{1}{2}}$ [81]. The delay in the formation of the metallic phase due to photoexcitation, which appears after the photoinduced melting of the excitonic insulator phase, has also been reported for Ta$_2$NiSe$_5$ and ascribed to this buildup time, $\tau_r$ [31,46]. In the case of photoexcitation, $\tau_r$ is approximately 0.4 ps and 0.1 ps for photocarriers 0.0016/Ni and 0.02/Ni, respectively. Similarly, for carriers generated by an electric field, a finite build-up time for carrier screening is likely to exist. In case the excitons are dissociated by an electric field, the spatial distribution of electron and hole carriers produced should be more biased, and $\tau_r$ can be expected to be longer than that in the case of photoexcitation.

As the electric field amplitude decreases, the number of excitons dissociated by quantum tunneling decreases rapidly. In this case, the energy position at which the reflectivity begins to increase in the $\Delta R(1\ \text{ps})$ spectrum is expected to move toward the lower-energy side. In the electric-field amplitude dependence of $\Delta R(1\ \text{ps})$ at 0.11 eV shown in Fig. 6(f), the negative $\Delta R$ signals in the range of 0.5–1.5 MV/cm can be understood as a response caused by the energy corresponding to the plasma frequency being below the probe energy of 0.11 eV.

# IV. DISCUSSION

## A. SIMULATIONS OF TRANSIENT REFLECTIVITY AND OPTICAL CONDUCTIVITY SPECTRA

As noted in Section III, the $\Delta R$ spectrum at 1 ps is caused by carrier generation and is attributable to the reduction and low-energy shift of the absorption peak characteristic of the exciton-condensed state and the appearance of a Drude response below 0.15 eV. To verify this interpretation, we simulated the $\Delta R$ spectrum. In this simulation, we assumed that the change in the complex dielectric constant $\Delta\tilde{\varepsilon}(\omega)$ induced by the terahertz electric field consisted of three components, as shown below:

$$\Delta\tilde{\varepsilon}(\omega) = \Delta\varepsilon_\infty + \Delta\tilde{\varepsilon}_{\text{ex}}(\omega) + \tilde{\varepsilon}_{\text{c}}(\omega). \tag{5}$$

On the right-hand side, the first term represents the change in the dielectric constant, $\varepsilon_\infty$, in the high frequency range. The second and third terms represent the changes in the transition around 0.4 eV inherent in the exciton-condensed state and the metallic response due to the generated carriers, respectively.

First, to describe the transition inherent in the exciton-condensed state in the second term, we used the analytical results of the linear and terahertz electro-reflectance spectra described in Section III. As mentioned in Subsection III-B, the steady-state $\sigma_1$ spectrum is considered to consist of $\sigma_{1\text{ex}}$ and $\sigma_{1\text{b}}$, reflecting the band-edge transition due to the exciton-condensed state and the interband transition at the higher energy side [Figs. 3(a) and 5(c)]. This section provides a more detailed analysis of the steady-state and transient spectra using the complex dielectric constant $\tilde{\varepsilon}(\omega)$. In the analysis, we also assume that the transition to the excited state $|1\rangle$ at the frequency $\omega_1$ inherent to the exciton-condensed state is expressed by $\tilde{\varepsilon}_{\text{ex}}(\omega, \omega_1)$ and the component of the interband transition at the higher energy is expressed by $\tilde{\varepsilon}_{\text{b}}(\omega)$. $\tilde{\varepsilon}_{\text{b}}(\omega)$ can be obtained by subtracting $\tilde{\varepsilon}_{\text{ex}}(\omega, \omega_1)$ from the experimentally measured dielectric spectrum $\tilde{\varepsilon}(\omega)$:

$$\tilde{\varepsilon}_{\text{b}}(\omega) = \tilde{\varepsilon}(\omega) - \tilde{\varepsilon}_{\text{ex}}(\omega, \omega_1). \tag{6}$$

Here, for simplicity, we assume that with a terahertz pulse, $\tilde{\varepsilon}_{\text{b}}(\omega)$ remains unchanged, whereas $\tilde{\varepsilon}_{\text{ex}}(\omega, \omega_1)$ varies by $\Delta\tilde{\varepsilon}_{\text{ex}}(\omega)$, which can be expressed in terms of the change in transition intensity $I$, $\Delta I$, and the lower-energy shift in central frequency $\omega_1$ of the transition, $\Delta\omega_1$, as follows:

$$\Delta\tilde{\varepsilon}_{\text{ex}}(\omega) = \left(1 + \frac{\Delta I}{I}\right)\tilde{\varepsilon}_{\text{ex}}(\omega, \omega_1 + \Delta\omega_1) - \tilde{\varepsilon}_{\text{ex}}(\omega, \omega_1) \tag{7}$$

The response of carriers in Eq. (5), $\tilde{\varepsilon}_{\text{c}}(\omega)$, can be described using the following simple Drude response,

$$\tilde{\varepsilon}_{\text{c}}(\omega) = \frac{i}{\omega/\gamma_{\text{D}}} \frac{\omega_{\text{p}}^2/\gamma_{\text{D}}^2}{1 - i\,\omega/\gamma_{\text{D}}}, \tag{8}$$

in which $\omega_{\text{p}}$ and $\gamma_{\text{D}}$ are the plasma frequency and damping constant, respectively.

As shown in Fig. 2(e), several absorption peaks due to phonons exist in the terahertz region of Ta$_2$NiSe$_5$, which should attenuate the terahertz pulse used for excitation as it propagates through the crystal. Therefore, the numbers of excited states and carriers decrease as the distance from the crystal surface increases in the depth direction. To incorporate this effect, we assume that the complex dielectric constant $\tilde{\varepsilon}(z, \omega)$ at a position $z$ from the crystal surface can be expressed as follows:

$$\tilde{\varepsilon}(z, \omega) = \tilde{\varepsilon}_{\text{tr}}(\omega)\exp\left(-\frac{z}{l}\right) + \tilde{\varepsilon}(\omega)\left(1 - \exp\left(-\frac{z}{l}\right)\right) \tag{9}$$

Here, the dielectric constant of a transient state, $\tilde{\varepsilon}_{\text{tr}}(\omega)$, is given by $\tilde{\varepsilon}_{\text{tr}}(\omega) = \tilde{\varepsilon}(\omega) + \Delta\tilde{\varepsilon}(\omega)$. Parameter $l$ represents a specific depth when the excited states are exponentially distributed.

To determine the unknown parameter $l$, we simulated the variation in a terahertz electric-field waveform propagating through the crystal based on the complex dielectric constant in the terahertz region reported previously [76]. Furthermore, the $z$-dependence of the number of carriers is obtained using the electric-field dependence of $\Delta R(1\,\mathrm{ps})$ shown in Fig. 6 (e), from which $l$ can be evaluated to be 3.7 μm. The details of the simulation of the electric-field waveform of the terahertz pulse inside the crystal, dependence of $\Delta R(1\,\mathrm{ps})$ on the effective electric field amplitude, and $z$-dependence of the number of carriers are reported in Appendix D.

Because the complex dielectric constant $\tilde{\varepsilon}(z,\omega)$ is a function of $z$, it is useful for the analysis of the transient reflectivity spectra, $R + \Delta R$, to use the multilayer model, in which a sample consists of many thin layers having different complex dielectric constants [82]. Using the multilayer model with $\tilde{\varepsilon}(z,\omega)$ and $l = 3.7\,\mathrm{μm}$, we obtain the $\Delta R(1\,\mathrm{ps})$ spectrum represented by the green line in Fig. 9(b), which reproduces well the experimental one (the red circles). In Fig. 9(a), the original $R$ spectrum is provided for comparison. The used parameters are listed in Table II, from which we can see that the intensity $I$ of the transition specific to the exciton-condensed state reduces as $\Delta I/I = -0.034$ and its energy decreases as $\hbar\Delta\omega_1 = -0.028$ eV.

Next, based on the fitting result, we extrapolate the $\Delta R(1\,\mathrm{ps})$ spectrum and perform its KK transformation to obtain the phase change, $\Delta\theta(1\,\mathrm{ps})$. To derive the change in the optical conductivity, $\Delta\sigma_1$, we also use the multilayer model. Here, $\Delta\sigma_1$ corresponds to $\Delta\tilde{\varepsilon}(\omega) = \tilde{\varepsilon}_{\mathrm{tr}}(\omega) - \tilde{\varepsilon}(\omega)$ in Eq. (7) and, therefore, is the change in optical conductivity at the crystal surface ($z = 0$). The derivation procedure of $\Delta\sigma_1$ is reported in Appendix E. The obtained $\Delta\sigma_1$ is shown by the red line in Fig. 9(d), together with the original $\sigma_1$ spectrum in Fig. 9(c), which is the same as Fig. 3(a). If the $\Delta\sigma_1$ spectrum is reasonably derived using the multilayer model, it should be reproduced using the parameters employed for the fitting analysis of $\Delta R(1\,\mathrm{ps})$ in

Table II. To verify this behavior, Fig. 9(d) plots the change in the optical conductivity $\sigma_{\text{ex}}$ associated with the transition to $|1\rangle$ by the orange dashed line and the Drude component arising from carrier generations by the blue dashed line, both of which are calculated using the fitting parameters in Table II. We also show the sum of these two components by the green line in the same figure, which reproduces the experimental $\Delta\sigma_1$ spectrum represented by the red line. This reproduction demonstrates that our analysis of the $\Delta\sigma_1$ spectrum is valid.

With the success of this simulation, the response of the excitonic insulator phase to a strong terahertz electric field can be interpreted as follows. Under the influence of the terahertz electric field, a portion of the initially condensed excitons dissociated, leading to the partial melting of the exciton condensation, as shown in the third panel of Fig. 7(a). The generated carriers behaved as free carriers, reducing the absorption specific to the original exciton condensate. Additionally, the screening effect due to carrier generation reduces the Coulomb attractive interactions within the remaining excitons, causing the absorption specific to the exciton-condensed state originally located at approximately 0.36 eV to shift to a lower energy, which is schematically illustrated in the fourth panel of Fig. 7(a).

Here, we discuss two important parameters derived from the analyses of the results in this study: the electron–hole correlation length and the density of the electric field-induced carriers. Using the threshold electric field, $E_{\text{th}}$, of the carrier generations, we can estimate the characteristic distance of the fluctuating electrons and holes near the bottom of the conduction band and near the top of the valence band, respectively, in the excitonic phase, which is hereafter called the correlation length $\xi$. The presence of strong phonon absorptions suppresses the electric-field amplitude of the terahertz pulse within the crystal via Fresnel reflection, whereas the large dielectric constant in the terahertz region effectively enhances the local electric field. Considering these effects, the effective value of the threshold electric field, $E_{\text{th}}$, was estimated to be approximately 2.37 MV/cm. The detail of the estimation is also reported in Appendix D. Using this $E_{th}$ value and the optical gap $\Delta_{gap} = 0.36$ eV in Ta$_2$NiSe$_5$, $\xi$ can be estimated to be approximately 7.6 Å from the relationship $\xi = \Delta_{gap}/(2eE_{th})$. Therefore, the effective distance of an electron and a hole in an exciton in the exciton-condensed state of Ta$_2$NiSe$_5$ is expected to be slightly smaller than $\xi = 7.6$ Å. A recent study based on ARPES demonstrated that the exciton radius $r_{ex} = 7.5$ Å [83]. Considering the density of an electron and a hole, the effective distance between an electron and a hole in an exciton is half that of $r_{ex}$, $r_{ex}/2 = 3.75$ Å. The obtained relation $\xi > r_{ex}/2$ supports the validity of our interpretation that the electrons (holes) in the exciton-condensed state tunnel into the conduction (valence) band due to the terahertz electric field.

Regarding the carrier density $N_c$, the simple Drude model gives the relation $\omega_p = \left(\frac{N_c e^2}{m^* \varepsilon_0}\right)^{\frac{1}{2}}$, where $e$ is the elementary charge and $m^*$ is the effective mass. The effective mass estimated from the band structure in the excitonic phase was $m^* = 5m_0$ ($m_0$: free-electron mass) [31]. In our experiments, the carriers generated by the terahertz electric field were expected to behave as free carriers because they exhibited a simple Drude response. The effective mass after the exciton-condensed-state melting was estimated to be $m^* = 0.37m_0$ by Tr-ARPES. Because the excitonic phase was not completely melted in our experiments, $m^*$ of the electric field-induced carriers can reasonably be assumed to be between $5m_0$ and $0.37m_0$. From these two values and $\hbar\omega_p = 0.90$ eV, we can consider that the carrier density would be a value between $\widetilde{N}_c = 0.3$ Ni$^{-1}$ and $0.02$ Ni$^{-1}$. As mentioned in the next section, the excitonic state is almost melted by the 1.55 eV excitation with an excitation photon density greater than 0.1 photon(ph)/Ni, which corresponds to $\widetilde{N}_c = 0.2$ Ni$^{-1}$, whereas only a small part is melted by the terahertz electric field. Therefore, we assume that $\widetilde{N}_c$ is much smaller than $0.3$ Ni$^{-1}$ and rather close to $0.02$ Ni$^{-1}$. Precise estimation of $\widetilde{N}_c$ remains a subject for future research.

## B. COMPARISON OF PHOTOINDUCED AND TERAHERTZ ELECTRIC-FIELD INDUCED ELECTRONIC STATE CHANGES

Finally, we compare the metallization of Ta$_2$NiSe$_5$ induced by the terahertz electric field reported here with that induced by optical excitation with a 1.55 eV pulse [46]. Recently, we reported the reflectivity change spectrum from the NIR to the MIR region when excited at 1.55 eV. The spectral changes of the reflectivity, $\Delta R_{\rm ph}(t)$, at $t = 0.2$ ps, $\Delta R_{\rm ph}(0.2\ {\rm ps})$, obtained with a significantly weak excitation with 0.0004 photons(ph)/Ni and strong excitation with 0.1 ph/Ni are shown in Figs. 9(e) and (f), respectively. With weak excitation [Fig. 9 (e)], $\Delta R_{\rm ph}(0.2\ {\rm ps})$ around the original reflectivity peak at 0.4 eV hardly decreases, in contrast to the case of the terahertz pulse excitation shown in Fig. 9(b). With decreasing photon energy below 0.4 eV, $\Delta R_{\rm ph}(0.2\ {\rm ps})$ increases, showing a peak structure around 0.3 eV and takes a finite value below 0.2 eV. On the other hand, with strong excitation [Fig. 9(g)], $\Delta R_{\rm ph}(0.2\ {\rm ps})$ monotonically increases with decreasing photon energy, suggesting that the metallic state is generated.

To obtain the spectral change of the optical conductivity, $\Delta\sigma_{1{\rm ph}}(t)$, at $t = 0.2$ ps [$\Delta\sigma_{1{\rm ph}}(0.2\ {\rm ps})$], which is derived from the $\Delta R_{\rm ph}(0.2\ {\rm ps})$ spectrum, we also adopt the multilayer model used in the analysis of the $\Delta\sigma_1$ spectrum obtained by the terahertz pulse excitation mentioned above. Because the absorption depth of the pump light with 1.55 eV is quite short (approximately 27 nm) and that of the probe light is longer, the multilayer model is also effective. The $\Delta\sigma_{1{\rm ph}}(0.2\ {\rm ps})$ spectra at 0.0004 ph/Ni and 0.1 ph/Ni are shown by the red lines in Figs. 9(g) and (h), respectively. The derivation of these spectra is presented in Appendix F. In Figs. 9(g) and (h), the $\sigma_1$ spectrum specific to the exciton-condensed state, $\sigma_{1{\rm ex}}$, is shown for comparison. In the weak excitation case (0.0004 ph/Ni), the $\Delta\sigma_{1{\rm ph}}(0.2\ {\rm ps})$ spectrum [the red line in Fig. 9(f)] shows that the optical conductivity around the original peak at 0.4 eV slightly decreases and the spectral weight around 0.3 eV increases. This spectral change resembles the $\Delta\sigma_1(1\text{ ps})$ spectrum induced by the terahertz electric field. Therefore, this spectrum was analyzed using Eq. (7), which considers the change in $\sigma_{1\text{ex}}$. Using Eq. (7), the $\Delta\sigma_{1\text{ph}}(0.2\text{ ps})$ spectrum above 0.25 eV is well reproduced, as shown by the green line. Appendix F presents the details of the analysis. The used parameters are $\Delta I/I = 0.010$ and $\hbar\Delta\omega_1 = -0.0039$ eV, which indicate the increase and lower energy shift of the $\sigma_{1\text{ex}}$ spectrum. Note that $\sigma_{1\text{ex}}$ increases unlike in the case of terahertz pulse excitation.

Besides this spectral change, a finite increase in $\Delta\sigma_1$ appears in the lower energy region. This component is ascribed to the photogenerated carriers. The electronic state change owing to this weak photoexcitation is illustrated in Fig. 7(b) and can be interpreted as follows. The electrons and holes generated by the interband transition at high-energy (1.55 eV) excitation immediately relaxed to the bottom of the conduction band and the top of the valence band, respectively. Subsequently, some of these electrons and holes form excitons [fourth panel in Fig. 7(b)], contributing to the increase in the spectral weight at the absorption edge, and the other part remains as unbounded carriers. The exciton-condensed state should be somewhat unstable compared to the original one because of the increase in the exciton number [fourth panel in Fig. 7(b)] and the generation of free carriers, although both increases are small. That is, the effective Coulomb attractive interaction within each exciton is slightly suppressed, resulting in a low-energy shift of $\sigma_{1\text{ex}}$ specific to the exciton-condensed state. This behavior resembles the phenomenon expected in optically excited semiconductors, in which the Bose-Einstein condensation of excitons becomes unstable in the case of a high density of excitons [84,85].

Looking at the $\Delta\sigma_{1\text{ph}}(0.2\text{ ps})$ spectrum in the case of the strong excitation with 0.1 ph/Ni [red line in Fig. 9(h)], the spectral weight around the original peak at 0.4 eV in the $\sigma_1$ spectrum is considerably decreased. In contrast, the spectral weights at lower energies are increased significantly. These spectral changes indicate a photoinduced insulator–metal transition. By scrutinizing this spectral change, a small shoulder structure is observed at approximately 0.32 eV, which is a signature of the low energy shift of the $\sigma_{1\text{ex}}$-component peak at 0.36 eV due to the exciton-condensed state. The $\sigma_1$ value of the original peak is approximately 4000 S/cm, whereas $\Delta\sigma_{1\text{ph}}(0.2\text{ ps})$ reaches –3300 S/cm. This finding indicates that the exciton-condensed state is almost completely melted, and the remaining excitonic states are quite small. This result is illustrated in Fig. 7(c) and can be interpreted as follows [46]. In the case of strong photoexcitation, many electrons and holes that behave as free carriers are generated, resulting in a Drude-like spectrum. These carriers render the excitonic phase unstable and melt the exciton-condensed state. Although the original peak specific to the exciton-condensed state does not completely disappear and a small part of the excitonic state may remain, the number of excitons is extremely small, as shown in the third panel in Fig. 7(c). Some of the previous studies on photoexcited Ta$_2$NiSe$_5$ have suggested that photoinduced metallization does not occur [30,39]. It may be ascribed to the difference in the excitation conditions because metallization by the 1.55 eV pulse excitation is driven only for excitation photon densities larger than 0.1 ph/Ni. Most experiments in which photoinduced metallization was not observed were performed at low excitation photon densities [30,39].

However, in the case of terahertz pulse excitation, as shown in Fig. 7(a), the excitons that form the original exciton-condensed state are dissociated by a strong electric field, and carriers are generated by the excitons. This exciton dissociations necessarily resulted in a decrease in the exciton number and a reduction in the spectral weight around 0.36 eV, $\sigma_{1\text{ex}}$. Simultaneously, this $\sigma_{1\text{ex}}$ spectrum is shifted to a lower energy owing to the suppression of Coulomb attractive interactions within the remaining excitons. Although the dissociated carriers are not excited to high energies, as in the case of optical excitation, they behave as free carriers, and a Drude response appears. Some previous studies have suggested that a structural change from monoclinic to orthorhombic occurs with strong 1.55 eV excitation [33,46], which occurs when the exciton photon density exceeds 0.1 ph/Ni according to Ref. [46]. As the magnitude of $\Delta\sigma_1$ below 0.4 eV due to the terahertz electric field shown in Fig. 9(d) is considerably smaller than that of $\Delta\sigma_{1ph}$ upon strong photoexcitation with a photon density of 0.1 ph/Ni shown in Fig. 9(h), no structural change is considered to occur. However, concluding whether a structural change occurs in the case of terahertz pulse excitation is difficult based on the experiments performed in this study. Clarifying the structural dynamics is a topic for future research.

## V. CONCLUSION

In this study, we investigated the response of Ta$_2$NiSe$_5$ to a terahertz pulse by measuring the transient reflectivity change spectra and examined the melting of the excitonic phase by dissociating excitons under a strong electric field. To elucidate the origin of the absorption at the band edge of Ta$_2$NiSe$_5$, we first attempted electro-reflectance spectroscopy using a terahertz pulse with a relatively low amplitude of 0.3 MV/cm. A response following the square of the electric field was observed near the reflectivity peak around 0.4 eV. With increasing photon energy, we observed positive, negative, and positive reflectivity changes around the peak, which sometimes appeared when an electric field was applied to the 1D Mott insulators. Therefore, we analyzed the obtained results using a three-level model considering a third-order nonlinear optical effect, which was applied to 1D Mott insulators and successfully reproduced the changes in the complex dielectric constant spectra derived from the experimentally obtained reflectivity change spectrum. From these analyses, we assigned the absorption at 0.36 eV observed at 10 K to the exciton-condensed state.

Next, we measured the reflectivity change spectrum upon irradiation with a more intense terahertz pulse having an amplitude of up to 1.65 MV/cm. When probed at 0.39 eV near the reflectivity peak, the reflectivity change at the time origin could be interpreted as a third-order nonlinear optical response. However, the magnitude of the reflectivity change 1 ps after excitation, when the electric field had completely disappeared, exhibited a threshold-like dependence on the electric-field amplitude. When the electric-field amplitude exceeded 1.5 MV/cm, the reflectivity at 1 ps monotonically increased toward lower energies, showing metallic behavior. These results demonstrate that the excitons are dissociated into carriers by a strong electric field via quantum tunneling. Moreover, at the same delay time of 1 ps, negative and positive reflectivity changes appeared near the original reflectivity peak around 0.4 eV with a decrease in the photon energy. These changes were reproduced by the intensity decrease and low-energy shift of the transition specific to the exciton-condensed state, which could be attributed to the generation of free carriers and screening of Coulomb interactions in residual excitons. These spectral changes were fundamentally different from those observed during photoexcitation with a photon energy of 1.55 eV, which induced higher-energy band-to-band transitions. In the case of relatively weak photoexcitation with a photon energy of 1.55 eV, the absorption specific to the exciton-condensed state at the band edge never decreased but rather was enhanced, accompanied by a slightly lower energy shift. Strong photoexcitation with a photon energy of 1.55 eV can melt the exciton-condensed state and change the system to a metallic state. These dissimilar responses can be explained by the differences in the carrier generation mechanisms between the two cases. The method of applying the terahertz pulse used in this study is expected to be effectively utilized for interpreting the optical spectra of various excitonic insulator materials and exploring excitonic insulator-to-metal transitions caused by the dissociation of excitons.

Acknowledgements

This work was partly supported by a Grant-in-Aid for Scientific Research from the Japan Society for the Promotion of Science (JSPS) (Project Numbers: JP20K03801, JP21H04988) and CREST (JPMJCR1661), Japan Science and Technology Agency. N.T. was supported by Support for Pioneering Research Initiated by Next Generation of Japan Science and Technology Agency (JST SPRING). M.Y. was supported by World-leading Innovative Graduate Study Program for Materials Research, Information, and Technology (MERIT-WINGS) of the University of Tokyo. Y.H was supported by the University Fellowship Program for Science and Technology Innovations of Japan Science and Technology Agency (WINGS-QSTEP).

# APPENDIX A: VERIFICATION OF POSSIBLE FRANZ-KELDYSH EFFECT IN TERAHERTZ ELECTRIC FIELD RESPONSE

Here, we consider the possibility of the Franz-Keldysh effect [77,78] as the origin of the spectral change in Figs. 5(b) and (d) produced by the terahertz electric field. This effect is caused by the penetration of the electron wave function into the bandgap due to the application of an electric field. Consequently, the absorption edge moves toward the lower-energy side, and an oscillating structure appears on the higher-energy side of the absorption edge, depending on the magnitude of the electric field. For 1D electronic systems, the change in the dielectric constant $\tilde{\varepsilon}$ due to the Franz-Keldish effect can be expressed as follows [86]:

$$\tilde{\varepsilon}(\omega, E_\mathrm{f}) = \varepsilon_\infty + CF\left(\frac{\hbar(\omega + i\varGamma_0)}{E_\mathrm{g}}, \frac{E_\mathrm{f}}{E_\mathrm{g}}\right) \tag{10}$$

$$F(x,y) = \frac{1}{x^2\sqrt{y}}\left\{Ai\left(\frac{1-x}{y}\right)Bi\left(\frac{1-x}{y}\right) + iAi^2\left(\frac{1-x}{y}\right) \right. \\ \left. + Ai\left(\frac{1+x}{y}\right)Bi\left(\frac{1+x}{y}\right) - iAi^2\left(\frac{1+x}{y}\right) - 2Ai\left(\frac{1}{y}\right)Bi\left(\frac{1}{y}\right)\right\} \tag{11}$$

Here, $E_\mathrm{g}$ is the energy of the band gap, $\varGamma_0$ is the inhomogeneous width of interband transitions, $\varepsilon_\infty$ is the high-frequency dielectric constant, and $C$ is a constant intrinsic to the material. $E_\mathrm{f}$ corresponds to the energy of the electric field and can be expressed as $(\hbar eE/2\mu)^{1/3}$ using the electric field strength $E$ and the reduced carrier mass, $\mu$. $Ai$ ($Bi$) represents the Airy function of the first (second) type.

First, using Eq. (10) for $E_\mathrm{f} = 0$, we simulate the optical conductivity spectrum, $\sigma_1 = \omega\varepsilon_0\varepsilon_2$, at 10 K in the absence of an electric field. By adopting the parameters of $E_\mathrm{g} = 0.37$ eV, $\varGamma_0 = 66$ meV, and $C = 296$, the $\sigma_1$ spectrum (red line) was almost completely reproduced, as shown by the blue line in Fig. 10(a). Next, we attempted to reproduce the $\Delta\tilde{\varepsilon}$ spectra shown in Figs. 5(d) and (e) using Eq. (11). When $\mu$ was set to $5m_0$ as reported in a previous study [31], the $\Delta\tilde{\varepsilon}$ spectra should be dominated by the electric field strength, $E$. In Figs. 10(b) and (c), the green lines show the $\Delta\tilde{\varepsilon}$ spectra at $E = 627$ kV/cm, which gives almost the same negative value of $\Delta\varepsilon_2$ experimentally evaluated at 0.37 eV. However, the spectral shapes of the simulated $\Delta\varepsilon_1$ and $\Delta\varepsilon_2$ values do not agree with those experimentally obtained (the red lines). Reproducing the $\Delta\varepsilon_1$ and $\Delta\varepsilon_2$ spectra using other $E$ values was also difficult. From these results, we can conclude that the experimental $\Delta\tilde{\varepsilon}$ spectra are difficult to explain by the Franz-Keldysh effect.

**APPENDIX B: ANALYSIS OF THIRD-ORDER NONLINEAR SUSCEPTIBILITY SPECTRA USING A THREE-LEVEL MODEL**

To evaluate the values of the third-order nonlinear susceptibility $\chi^{(3)}$, estimating an effective magnitude of the terahertz electric field inside the crystal is necessary. Because $\varepsilon_1$ is quite large in the terahertz region in Ta$_2$NiSe$_5$, we should consider the local electric field, $E_\mathrm{L}$. Here, we set $\varepsilon_1/\varepsilon_0$ to 80, which is its approximate average value in the terahertz region [76]. This value of $\varepsilon_1$ enhances the local electric field via the equation $E_\mathrm{L} = E + \frac{P}{3\varepsilon_0} = \left(\frac{2}{3} + \frac{\varepsilon_1}{3\varepsilon_0}\right)E$, where $E$ is the external electric field, $P$ is the polarization, and $\frac{P}{3\varepsilon_0}$ is an additional electric field. We also consider the Fresnel loss, which reduces the electric field inside the crystal to 20% of the external electric field. In this case, $E_\mathrm{THz}(0) = 0.316$ MV/cm in the free space gives the effective electric field $E_\mathrm{eff}$ as $E_\mathrm{eff} = \left(\frac{2}{3} + \frac{\varepsilon_1}{3\varepsilon_0}\right) \times 0.2 E_\mathrm{THz}(0) = 5.47 \times E_\mathrm{THz}(0) = 1.73$ MV/cm.

The third-order nonlinear susceptibility $\chi^{(3)}(-\omega_\sigma; \omega_i, \omega_j, \omega_k)$ is generally expressed based on perturbation theory as follows [87].

$$\chi^{(3)}(-\omega_\sigma; \omega_i, \omega_j, \omega_k)$$

$$= \frac{Ne^4}{6\varepsilon_0 \hbar^3} \wp \sum_{(a,b,c)} \left[ \frac{\langle 0|x|a\rangle\langle a|x|b\rangle\langle b|x|c\rangle\langle c|x|0\rangle}{(\omega_a - \omega_\sigma - i\gamma_a)(\omega_b - \omega_j - \omega_k - i\gamma_b)(\omega_c - \omega_k - i\gamma_c)} \right.$$

$$+ \frac{\langle 0|x|a\rangle\langle a|x|b\rangle\langle b|x|c\rangle\langle c|x|0\rangle}{(\omega_a + \omega_i + i\gamma_a)(\omega_b - \omega_j - \omega_k - i\gamma_b)(\omega_c - \omega_k - i\gamma_c)} \quad (12)$$

$$+ \frac{\langle 0|x|a\rangle\langle a|x|b\rangle\langle b|x|c\rangle\langle c|x|0\rangle}{(\omega_a + \omega_i + i\gamma_a)(\omega_b + \omega_i + \omega_j + i\gamma_b)(\omega_c - \omega_k - i\gamma_c)}$$

$$\left. + \frac{\langle 0|x|a\rangle\langle a|x|b\rangle\langle b|x|c\rangle\langle c|x|0\rangle}{(\omega_a + \omega_i + i\gamma_a)(\omega_b + \omega_i + \omega_j + i\gamma_b)(\omega_c + \omega_\sigma + i\gamma_c)} \right],$$

where $|0\rangle$ denotes ground state; $|a\rangle$, $|b\rangle$, and $|c\rangle$ represent the excited states; $\omega_i$ and $\gamma_i$ are the frequency and damping constant of state $|i\rangle$ ($i = a, b,$ or $c$); and $\langle i|x|j\rangle$ is the transition dipole moment between states $|i\rangle$ and $|j\rangle$. The relation $\omega_\sigma = \omega_i + \omega_j + \omega_k$ stands. $\wp$ denotes taking the sum of permutations of $(\omega_i, \omega_j, \omega_k)$. $N$ is the number of atoms per unit volume.

In the three-level model, the ground state $|0\rangle$, one-photon-allowed excited state with odd parity $|1\rangle$, and one-photon-forbidden excited state with even parity $|2\rangle$ are considered. In this model, the third-order nonlinear susceptibility $\chi^{(3)}(-\omega_\sigma; \omega_i, \omega_j, \omega_k)$ can be expressed as follows:

$$\chi^{(3)}(-\omega_\sigma; \omega_i, \omega_j, \omega_k)$$

$$= \frac{Ne^4}{6\varepsilon_0 \hbar^3} \wp \left[ \frac{\langle 0|x|1\rangle\langle 1|x|2\rangle\langle 2|x|1\rangle\langle 1|x|0\rangle}{(\omega_1 - \omega_\sigma - i\gamma_1)(\omega_2 - \omega_j - \omega_k - i\gamma_2)(\omega_1 - \omega_k - i\gamma_1)} \right.$$

$$+ \frac{\langle 0|x|1\rangle\langle 1|x|2\rangle\langle 2|x|1\rangle\langle 1|x|0\rangle}{(\omega_1 + \omega_i + i\gamma_1)(\omega_2 - \omega_j - \omega_k - i\gamma_2)(\omega_1 - \omega_k - i\gamma_1)} \quad (13)$$

$$+ \frac{\langle 0|x|1\rangle\langle 1|x|2\rangle\langle 2|x|1\rangle\langle 1|x|0\rangle}{(\omega_1 + \omega_i + i\gamma_1)(\omega_2 + \omega_i + \omega_j + i\gamma_2)(\omega_1 - \omega_k - i\gamma_1)}$$

$$\left. + \frac{\langle 0|x|1\rangle\langle 1|x|2\rangle\langle 2|x|1\rangle\langle 1|x|0\rangle}{(\omega_1 + \omega_i + i\gamma_1)(\omega_2 + \omega_i + \omega_j + i\gamma_2)(\omega_1 + \omega_\sigma + i\gamma_1)} \right]$$

In the above equation, we neglect terms for $(a, b, c) = (0,1,0)$ because their contributions are sufficiently small. Using Eq. (12), we obtain the following expression for $\chi^{(3)}(-\omega; 0,0, \omega)$ dominating terahertz electro-reflectance spectra:

$$\chi^{(3)}(-\omega; 0,0, \omega) = \frac{Ne^4}{3\varepsilon_0 \hbar^3} \langle 0|x|1\rangle\langle 1|x|2\rangle\langle 2|x|1\rangle\langle 1|x|0\rangle$$

$$\times \left[\frac{1}{(\omega_1 - \omega - i\gamma_1)(\omega_2 - i\gamma_2)(\omega_1 - i\gamma_1)}\right.$$

$$+ \frac{1}{(\omega_1 + \omega + i\gamma_1)(\omega_2 - i\gamma_2)(\omega_1 - i\gamma_1)}$$

$$+ \frac{1}{(\omega_1 + \omega + i\gamma_1)(\omega_2 + \omega + i\gamma_2)(\omega_1 - i\gamma_1)}$$

$$+ \frac{1}{(\omega_1 + \omega + i\gamma_1)^2(\omega_2 + \omega + i\gamma_2)}$$

$$+ \frac{1}{(\omega_1 - \omega - i\gamma_1)(\omega_2 - \omega - i\gamma_2)(\omega_1 - i\gamma_1)}$$

$$+ \frac{1}{(\omega_1 + i\gamma_1)(\omega_2 - \omega - i\gamma_2)(\omega_1 - i\gamma_1)} \quad (14)$$

$$+ \frac{1}{(\omega_1 + i\gamma_1)(\omega_2 + \omega + i\gamma_2)(\omega_1 - i\gamma_1)}$$

$$+ \frac{1}{(\omega_1 + i\gamma_1)(\omega_2 + \omega + i\gamma_2)(\omega_1 + \omega + i\gamma_1)}$$

$$+ \frac{1}{(\omega_1 - \omega - i\gamma_1)^2(\omega_2 - \omega - i\gamma_2)}$$

$$+ \frac{1}{(\omega_1 + i\gamma_1)(\omega_2 - \omega - i\gamma_2)(\omega_1 - \omega - i\gamma_1)}$$

$$+ \frac{1}{(\omega_1 + i\gamma_1)(\omega_2 + i\gamma_2)(\omega_1 - \omega - i\gamma_1)}$$

$$\left.+ \frac{1}{(\omega_1 + i\gamma_1)(\omega_2 + i\gamma_2)(\omega_1 + \omega + i\gamma_1)}\right]$$

We used this expression in the fitting analysis of the experimentally obtained $\chi^{(3)}(-\omega;0,0,\omega)$ spectrum.

**APPENDIX C: ESTIMATION OF TEMPERATURE INCREASE BY INTENSE TERAHERTZ PULSE EXCITATION**

The intensity of the incident terahertz pulse with an amplitude of 1.65 MV/cm was evaluated to be $I_{\text{in}} = 0.21 \text{ mJ/cm}^2$ when the reflection loss was considered. As the full width at half-maximum of the incident pulse was approximately 140 µm, the energy applied to the system was approximately 32 nJ. Using the previously reported specific heat $C(T)$ of Ta$_2$NiSe$_5$ [26], the energy, $W$, required to increase the temperature of the system from $T_0 = 10$ K to $T$ can be expressed as

$$W = V_{\text{pump}} \int_{T_0}^{T} C(T')dT' \qquad (15)$$

$V_{\text{pump}}$ is the volume excited by the terahertz pulse. The absorption coefficient $\alpha$ of the terahertz pulse depends on the frequency because strong absorptions due to phonons exist in the terahertz region [Fig. 2(f)]. At 4 THz, where $\alpha$ takes a maximum as shown in Fig. 2(f), the length, $l_{\text{p}} \left( = \frac{1}{\alpha} \right)$, from the crystal surface at which the intensity of the terahertz pulse is reduced to 1/e (e: Napier number) is approximately 0.5 µm. Assuming that the energy of the terahertz pulse is absorbed up to $l_{\text{p}}$ in the depth direction from the crystal surface following the corresponding $\alpha$ value and using Eq. (15), the temperature, $T$, after the excitation of the system is estimated to be 36.7 K, which is the maximum increase in temperature. This value is close to 60 K, which was used to calculate the temperature differential spectrum indicated by the orange line in Fig. 8(c). From this result, the change in reflectivity at 7 ps can be considered to be due mainly to the temperature increase caused by the terahertz pulse. If we use as $l_{\text{p}}$ the characteristic length, $l = 3.7$ μm, which is the depth for the presence of electric-field induced carriers as estimated in Appendix D, and the corresponding $\alpha(= 1/l)$ value, the temperature increase in the region up to $l = 3.7$ μm from the crystal surface can be estimated to be approximately 19.8 K, whereas the interpretation of the reflectivity change at 7 ps is essentially the same.

**APPENDIX D: DEPENDENCES OF ELECTRIC FIELD WAVEFORM AND CARRIER NUMBER ON THE DISTANCE FROM CRYSTAL SURFACE**

The change in the electric field waveform of the terahertz pulse inside the crystal was estimated as follows. Figure 11(a) shows the electric field waveform, $E_{\text{in}}(t)$, of the incident terahertz pulse. The power spectrum, $|\tilde{E}_{\text{in}}(\omega)|^2$, obtained by the Fourier transformation of $E_{\text{in}}(t)$ is shown by the red line in Fig. 11(b) and is identical to the red line in Fig. 2(c). Considering the Fresnel reflection at the crystal surface and absorption inside the crystal, the Fourier component $\tilde{E}_{\text{THz}}(z, \omega)$ of the terahertz electric field at a depth $z$ from the crystal surface can be expressed as follows:

$$\tilde{E}_{\text{THz}}(z, \omega) = \frac{2}{\tilde{n}(\omega) + 1} \exp\left(i \frac{(\tilde{n}(\omega) - 1)\omega z}{c}\right) \tilde{E}_{\text{in}}(\omega) \quad (16)$$

Here, we neglect nonlinear absorption due to carrier generation and consider only linear absorption.

The calculated power spectrum of the terahertz pulses at position $z$, $|\tilde{E}_{\text{THz}}(z, \omega)|^2$, is shown in Fig. 11(c). In this calculation, the results reported in Ref. [76] for the complex refractive index $\tilde{n}(\omega)$ in the terahertz region were employed. The $|\tilde{E}_{\text{THz}}(z, \omega)|^2$ spectra are represented by contour lines in Fig. 11(d). The Fresnel reflection at the crystal surface causes several dip structures, as shown in Fig. 11(c), which depend on the refractive index dispersions of the IR-active phonons [Fig. 2(d)]. At $z \geq 6.6$ μm, the high-frequency components of the electric field above 3.2 THz almost disappear due to the strong linear absorptions by the IR active phonons. Furthermore, we performed the inverse Fourier transform of these spectra and obtained the electric field waveforms of the terahertz pulse at position $z$, which are presented in Fig. 11(e). Considering the local electric field amplitude, $E_\text{L}$, mentioned in Appendix B, the magnitudes of the electric fields were determined as shown along the vertical axis in Fig. 11(e).

As described in Subsection III-C, the threshold electric field for the carrier generation, $E_\text{th}$, was estimated to be 0.8 MV/cm using the electric-field amplitudes of the terahertz pulses in the free space. However, the value inside the crystal should be used as the electric-field amplitude. Considering the Fresnel reflection and local electric field, as mentioned in Appendix B, and using the electric field value at 5 μm, which corresponds approximately to the absorption depth of the terahertz pulse, the effective electric field amplitude of the terahertz pulse is 2.96 times as large as that in free space, and $E_\text{th}$ can be estimated to be approximately 2.37 MV/cm.

Next, we estimated the distribution of the carriers induced by the terahertz electric field in the crystal. The red line in Fig. 11(f) shows the change in the maximum value $E_\text{peak}(z)$ of the terahertz electric field waveform $E_\text{THz}(z,t)$. $E_\text{peak}(z)$ monotonically decreases at $z < 6.6$ μm, but it starts to increase at $z \sim 6.6$ μm and then begins to decrease again at $z \sim 11$ μm. At $z > 6$ μm, the high-frequency components above 3.2 THz have almost disappeared [Fig. 11(c)] and the decrease in $E_\text{peak}(z)$ is small [Fig. 11(f)]. In addition, due to the narrowing of the spectrum, the electric field of the terahertz pulse tends to have multiple peaks for $z \geq 6.6$ μm, as shown in Fig. 11(e). Because the carrier generation probability $\Gamma$ due to quantum tunneling by the terahertz electric field can be expressed as $\Gamma \propto E_\text{THz} \exp(-\pi E_\text{th}/E_\text{THz})$, $\Gamma$ at position $z$, $\Gamma(z)$, can be expressed as follows:

$$\frac{\Gamma(z)}{\Gamma(0)} = \frac{E_\text{peak}(z)}{E_\text{peak}(0)} \exp\left[-\pi E_\text{th}\left(\frac{1}{E_\text{peak}(z)} - \frac{1}{E_\text{peak}(0)}\right)\right] \qquad (17)$$

For the threshold electric field, $E_{\text{th}}$, we use the corrected value of $E_{\text{th}} = 2.37 \, \text{MV/cm}$ as mentioned above. Consequently, $\Gamma(z)$ was obtained as shown by the blue line in Fig. 11(f). We fit this blue line with a single exponential function $\exp(-z/l)$, which approximately reproduces the z-dependent feature of the blue line, that is, $\Gamma(z)$, as shown by the green line in Fig. 11(f). From this analysis, the characteristic length $l$ for carrier generation was estimated to be approximately 3.7 μm.

## APPENDIX E: METHOD OF TRANSIENT OPTICAL CONDUCTIVITY SPECTRUM DERIVATION FROM TERAHERTZ ELECTRIC-FIELD INDUCED REFLECTIVITY CHANGES USING MULTILAYER MODEL

To apply the KK transformation to the $\Delta R$ spectrum, the $\Delta R$ data were extrapolated as follows. Below the lower energy bound of the measured region, 0.094 eV, the $\Delta R$ data were extrapolated using the simulation results [green line in Fig. 9(b)] described in Section IV. Above the higher energy bound of the measured region, 0.77 eV, the $\Delta R$ data were extrapolated using a straight line so that $\Delta R$ reached zero at 0.9 eV and $\Delta R = 0$ was assumed at higher energies. The KK transformation was applied to the $(R + \Delta R)$ spectrum, which was obtained by adding the steady-state reflectivity $R$ spectrum to the $\Delta R$ spectrum extrapolated as mentioned above, and the transient phase $(\theta + \Delta\theta)$ spectrum was calculated. Here, $\theta$ is a phase of the electric field of the reflected light in the steady state and $\Delta\theta$ is its photoinduced change.

Next, the multilayer model was used to obtain the spectrum of transient optical conductivity $\sigma_1 + \Delta\sigma_1$. In this multilayer model, the crystal is considered to consist of multilayers with exponentially varying complex dielectric constants, $\tilde{\varepsilon}(z, \omega)$, as in Eq. (9), where $l = 3.7$ μm determines the rate of change in $\tilde{\varepsilon}(z, \omega)$. The real and imaginary parts of the spectrum of the complex permittivity, $\tilde{\varepsilon}_{\text{tr}}(\omega)$, in Eq. (9) can be determined so as to reproduce $R + \Delta R$ and $\theta + \Delta\theta$. From the obtained $\tilde{\varepsilon}_{\text{tr}}(\omega)$ spectrum, the $\sigma_1 + \Delta\sigma_1$ spectrum at the crystal surface ($z = 0$) can be calculated. By substituting the steady-state $\sigma_1$ spectrum from the $(\sigma_1 + \Delta\sigma_1)$ spectrum, the $\Delta\sigma_1$ spectrum can be obtained as shown by the red line in Fig. 9(d).

**APPENDIX F: METHOD OF TRANSIENT OPTICAL CONDUCTIVITY SPECTRUM DERIVATION FROM PHOTO-INDUCED REFLECTIVITY CHANGES USING MULTILAYER MODEL**

This section describes the methods utilized to derive the spectral change of optical conductivity $\sigma_1$, $\Delta\sigma_{1\mathrm{ph}}$, from the spectrum of the reflectivity change $\Delta R_{\mathrm{ph}}$ obtained by the 1.55 eV excitation previously reported and to simulate the $\Delta\sigma_{1\mathrm{ph}}$ spectrum. First, the phase change, $\Delta\theta_{\mathrm{ph}}$, of the reflected light was calculated by using the KK transform from the $\Delta R_{\mathrm{ph}}$ spectrum. For the weak-excitation case (0.0004 ph/Ni), $\Delta R_{\mathrm{ph}}$ was very small for all measured photon energy regions. Therefore, as in the previous study [46], the $\Delta\theta_{\mathrm{ph}}$ spectrum was calculated directly from the $\Delta R_{\mathrm{ph}}$ spectrum using the following equation:

$$\Delta\theta_{\mathrm{ph}} = \frac{\omega}{\pi}\mathcal{P}\int_0^\infty \frac{\Delta R_{\mathrm{ph}}(\omega_1)}{R}\frac{1}{\omega_1^2 - \omega^2}d\omega_1, \tag{18}$$

where $\mathcal{P}$ denotes Cauchy's principal value. Below the lower-energy bound of the measured region (0.138 eV), $\Delta R_{\mathrm{ph}}$ was extrapolated as a constant value equal to $\Delta R_{\mathrm{ph}}$ at 0.138 eV. Above the higher-energy bound of the measured region, 1.39 eV, $\Delta R_{\mathrm{ph}}$ was extrapolated to be 0. $R_{\mathrm{ph}} + \Delta R_{\mathrm{ph}}$ and $\theta + \Delta\theta_{\mathrm{ph}}$ were obtained by adding $\Delta R_{\mathrm{ph}}$ and $\Delta\theta_{\mathrm{ph}}$ to $R$ and $\theta$ in the steady state.

For the strong excitation case (0.1 ph/Ni), the previous study extrapolated the low-energy side of the $R_{\mathrm{ph}} + \Delta R_{\mathrm{ph}}$ spectrum using the Hagen-Rubens relation. However, this extrapolation is not exactly appropriate when using the multilayer model, because the absorption depth $l$ of the pump light at 1.55 eV and 27 nm is quite short, i.e., the region in which the photoexcited states are generated is very thin. Therefore, in this study, below the lower-energy bound of the measured region (0.138 eV), $\Delta R_{\text{ph}}$ was extrapolated as a constant value equal to $\Delta R_{\text{ph}}$ at 0.138 eV. Above the higher energy bound of the measured region, 1.39 eV, $\Delta R_{\text{ph}}$ was extrapolated as a constant value of $\Delta R_{\text{ph}}$ at 1.39 eV up to 13 eV and as $\Delta R_{\text{ph}} \propto \omega^{-4}$ above 13 eV.

The derivation of the optical conductivity change, $\Delta\sigma_{1\text{ph}}$, at the crystal surface ($z = 0$) was conducted using the same multilayer model ($l = 27$ nm) that was utilized for the derivation of $\Delta\sigma_1$ due to the terahertz pulse excitation, which is detailed in Appendix E. The obtained $\Delta\sigma_{1\text{ph}}$ spectrum is slightly different from that previously reported in Ref. [46] because Ref. [46] did not use the multilayer model and assumed $l = \infty$ in Eq. (7). Note that the $\Delta\sigma_{1\text{ph}}$ spectrum obtained in the present study using the multilayer model is more accurate.

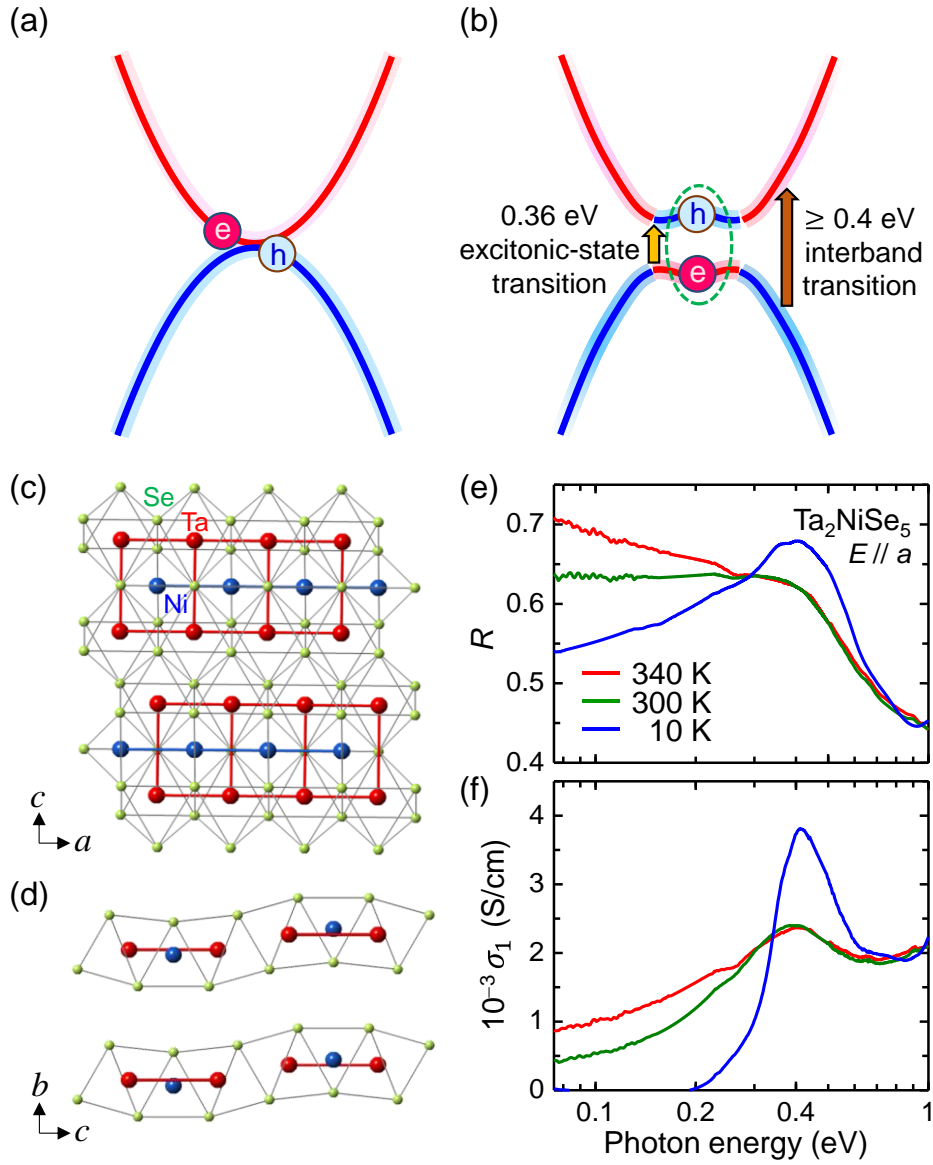

Fig. 1 (a,b) Schematic band structures of (a) a semimetal and (b) an excitonic insulator. (c,d) Crystal structure of $Ta_2NiSe_5$ viewed along the *b*-axis (c) and along the *a*-axis (d). (e) Polarized reflectivity ($R$) spectra with electric fields $E$ parallel (//) to the *a* axis in $Ta_2NiSe_5$. (f) Optical conductivity ($\sigma_1$) spectra deduced from the $R$ spectra in (e) using the KK transformation.

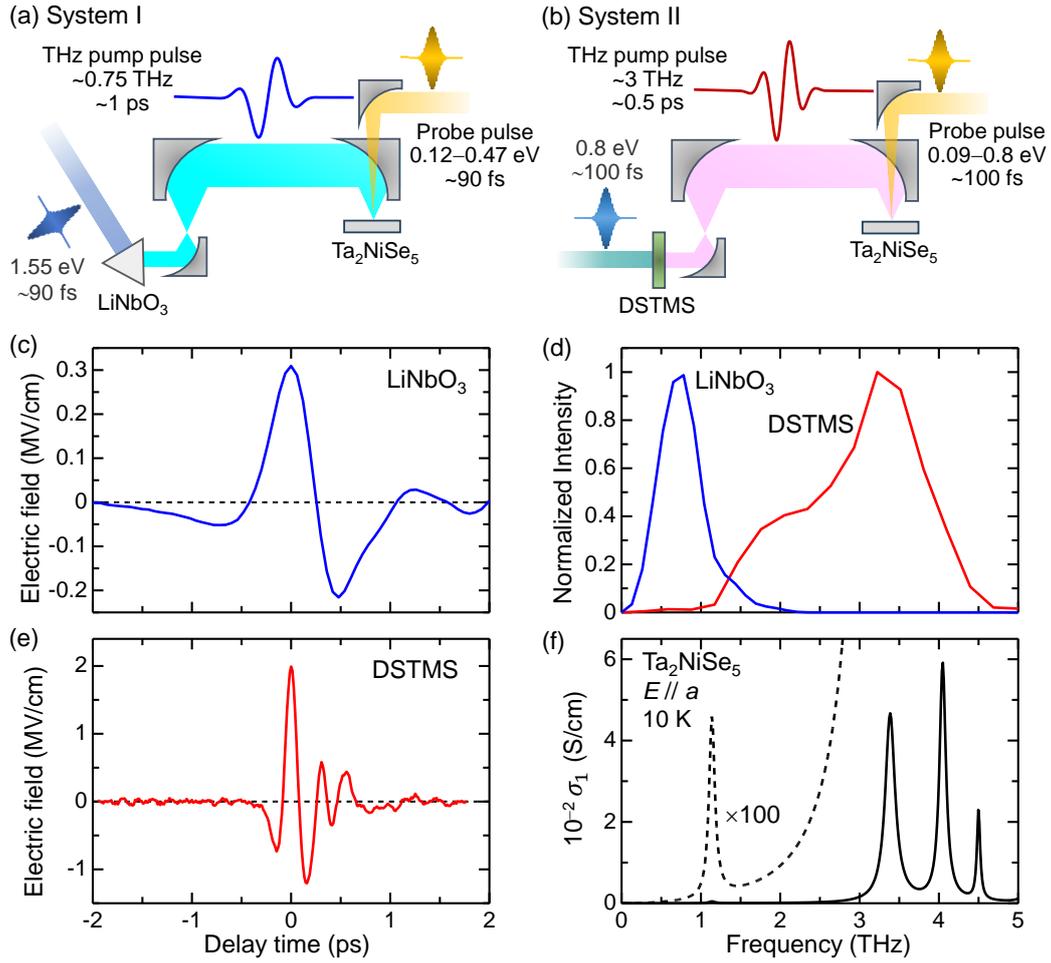

Fig. 2 (a,b) Schematics of terahertz pump optical reflection probe spectroscopy measurements. (a) system I and (b) system II use nonlinear optical crystals of LiNbO$_3$ and DSTMS to generate terahertz pulses, respectively. (c,d) A typical electric field waveform of a terahertz pulse generated using (c) LiNbO$_3$ and (d) DSTMS. (e) Intensity spectra of terahertz pulses generated using LiNbO$_3$ and DSTMS. (f) Optical conductivity ($\sigma_1$) spectrum at 10 K in Ta$_2$NiSe$_5$ reproduced by the parameters reported in ref. [76].

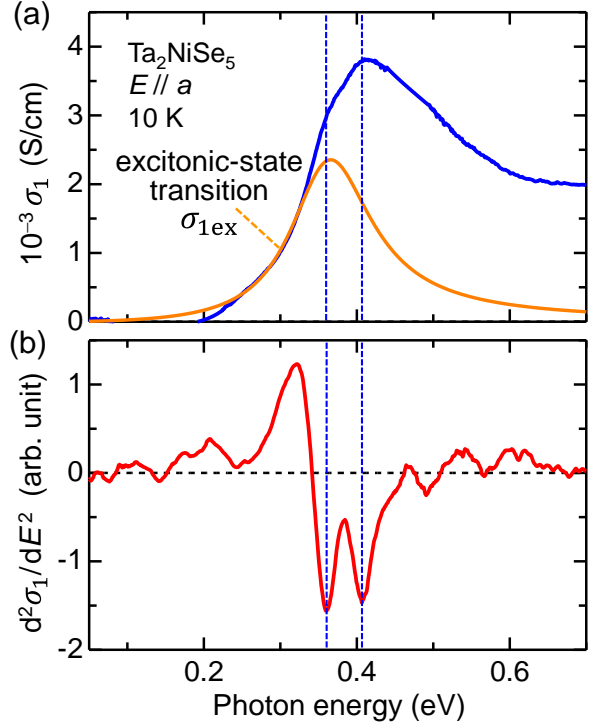

Fig. 3　Analyses of optical conductivity ($\sigma_1$) spectrum at 10 K in Ta$_2$NiSe$_5$. (a) The $\sigma_1$ spectrum for $E // a$ (the blue line) at 10 K and the component of the transition specific to the exciton-condensed state, $\sigma_{1\mathrm{ex}}$ (the orange line). (b) The second energy derivative of the $\sigma_1$ spectrum [the bule line in (a)]. The vertical blue broken lines show the energy positions of peak structures in the $\sigma_1$ spectrum in (a).

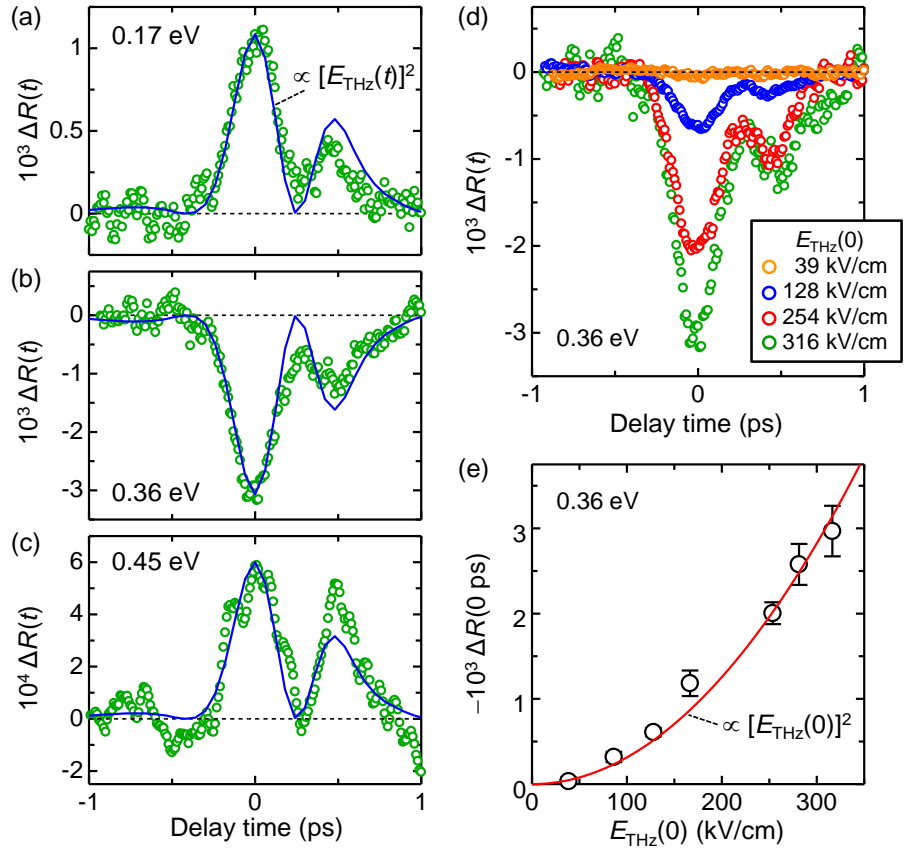

Fig. 4 (a-c) Time characteristics of terahertz electric-field induced reflectivity changes, $\Delta R(t)$, at 10 K for the probe energy of (a) 0.17 eV, (b) 0.36 eV, and (c) 0.45 eV in $Ta_2NiSe_5$. Electric fields of terahertz pump and MIR probe pulses are both parallel to the $a$ axis. The blue lines show time characteristics of $[E_{THz}(t)]^2$ in arbitrary units. (d) Time characteristics of $\Delta R$ for several typical electric field amplitudes of terahertz pump pulse, $E_{THz}(0)$, at 10 K. (e) $E_{THz}(0)$ dependence of $-\Delta R(0 \text{ ps})$ at 0.36 eV (10 K). The red line is proportional to $[E_{THz}(0)]^2$.

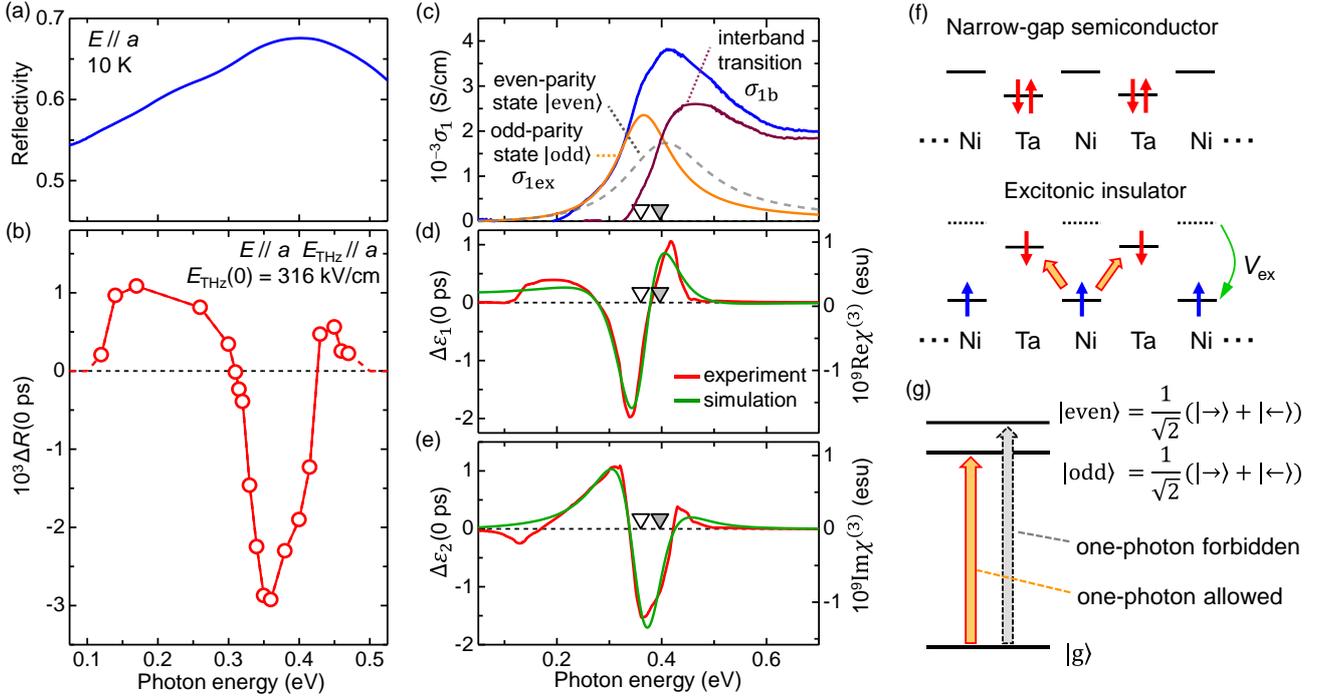

Fig. 5  Electro-reflectance spectrum at 10 K and its analysis in Ta$_2$NiSe$_5$. (a) The $R$ spectrum for $E \parallel a$ at 10 K and (b) its electric-field induced change, $\Delta R(0\,\mathrm{ps})$, for $E_{\mathrm{THz}}(0) = 316$ KV/cm. (c) The $\sigma_1$ spectrum (the blue line), the $\sigma_1$ component of the transition to odd-parity excited state $|\mathrm{odd}\rangle$, $\sigma_{1\mathrm{ex}}$ (the orange line), and the $\sigma_1$ component of the interband transition, $\sigma_{1\mathrm{b}}$ (the brown line). The gray line shows the even-parity excited state $|\mathrm{even}\rangle$ in an arbitrary unit for comparison. The orange line is the same as that shown in Fig. 3(a). (d,e) The spectra of (d) $\Delta\varepsilon_1(0\,\mathrm{ps})$ and (e) $\Delta\varepsilon_2(0\,\mathrm{ps})$ (the red lines) derived from the $\Delta R(0\,\mathrm{ps})$ spectrum shown in (b) using the KK transformation. The green lines show the fitting curves based upon the three-level model. (f) Schematic energy levels of Ta$_2$NiSe$_5$ shown by a hole picture. (g) Schematic of two excited states, $|\mathrm{odd}\rangle$ and $|\mathrm{even}\rangle$, and ground state $|g\rangle$ dominating the response to the electric field.

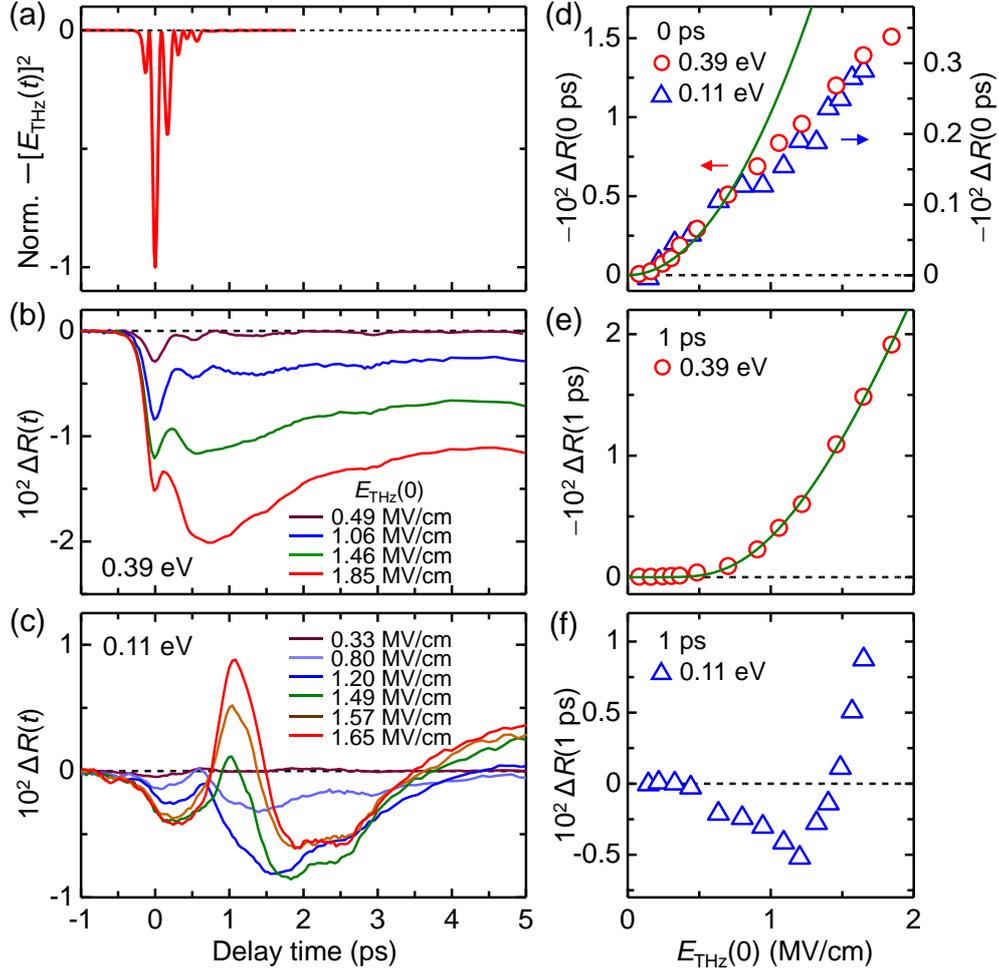

Fig. 6 Dependence of reflectivity changes $\Delta R(t)$ on the electric-field amplitude $E_{\mathrm{THz}}(0)$ of the terahertz pump pulse at 10 K. (a) Time characteristic of the square of the terahertz electric field, $[E_{\mathrm{THz}}(t)]^2$. (b,c) $E_{\mathrm{THz}}(0)$ dependence of time characteristics of $\Delta R(t)$ at (b) 0.39 eV and 0.11 eV. (d-f) $E_{\mathrm{THz}}(0)$ dependence of (d) $-\Delta R(0\,\mathrm{ps})$ at 0.39 eV and 0.11 eV, (e) $-\Delta R(1\,\mathrm{ps})$ at 0.39 eV, and $-\Delta R(1\,\mathrm{ps})$ at 0.11 eV.

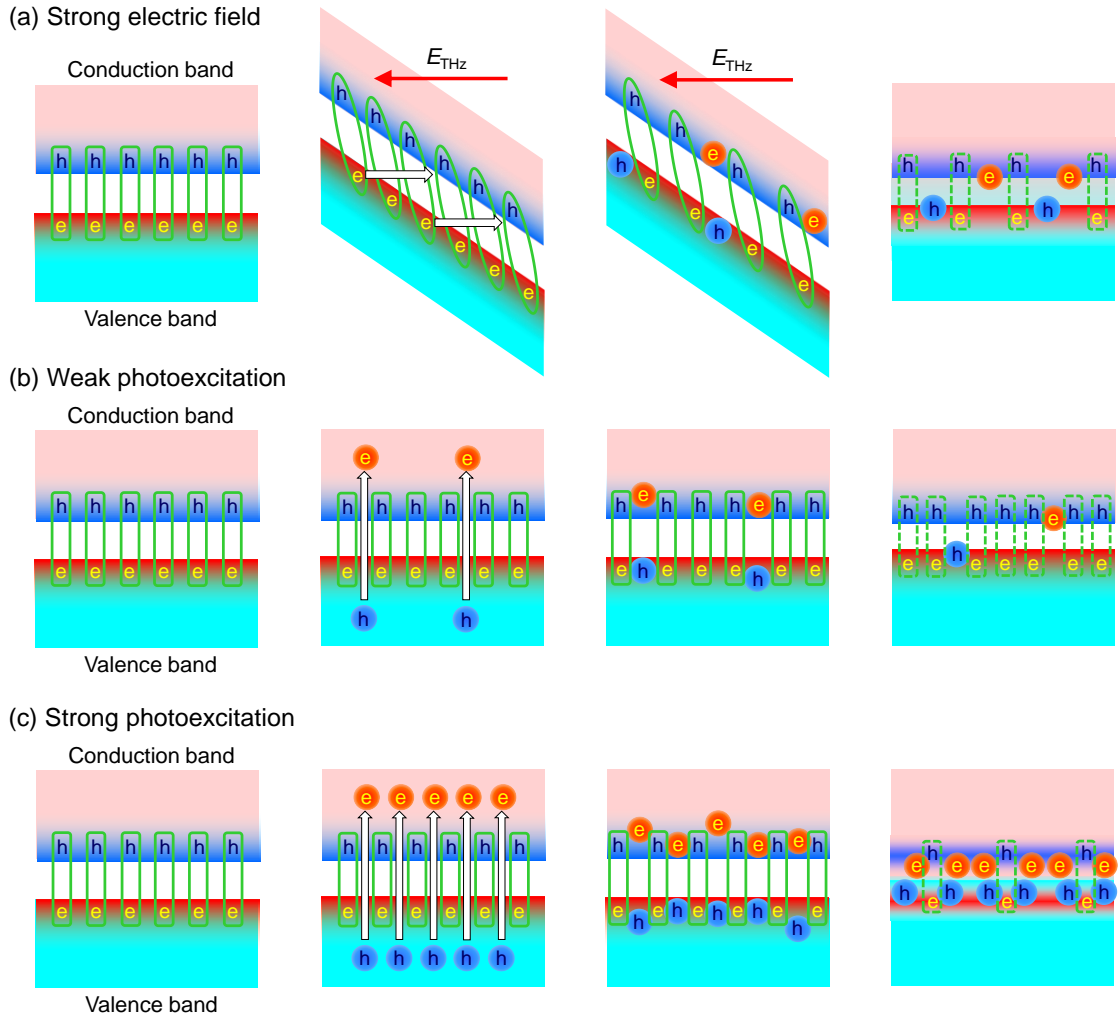

Fig. 7  Schematics of electronic-state changes by (a) a strong electric field $E_{\text{THz}}(0)$, (b) a weak photoexcitation with 1.55 eV, and (c) a strong photoexcitation with 1.55 eV in Ta$_2$NiSe$_5$. e and h denote an electron and a hole, respectively. The red and blue areas show the states composed of Ta 5d orbitals and those composed of a hybridization of Ni 3d orbitals and Se 4p orbitals, respectively. The electron-hole pairs enclosed by the green solid (broken) lines represent strongly (weakly) bound excitons. The red and blue circles show free electrons and holes, respectively.

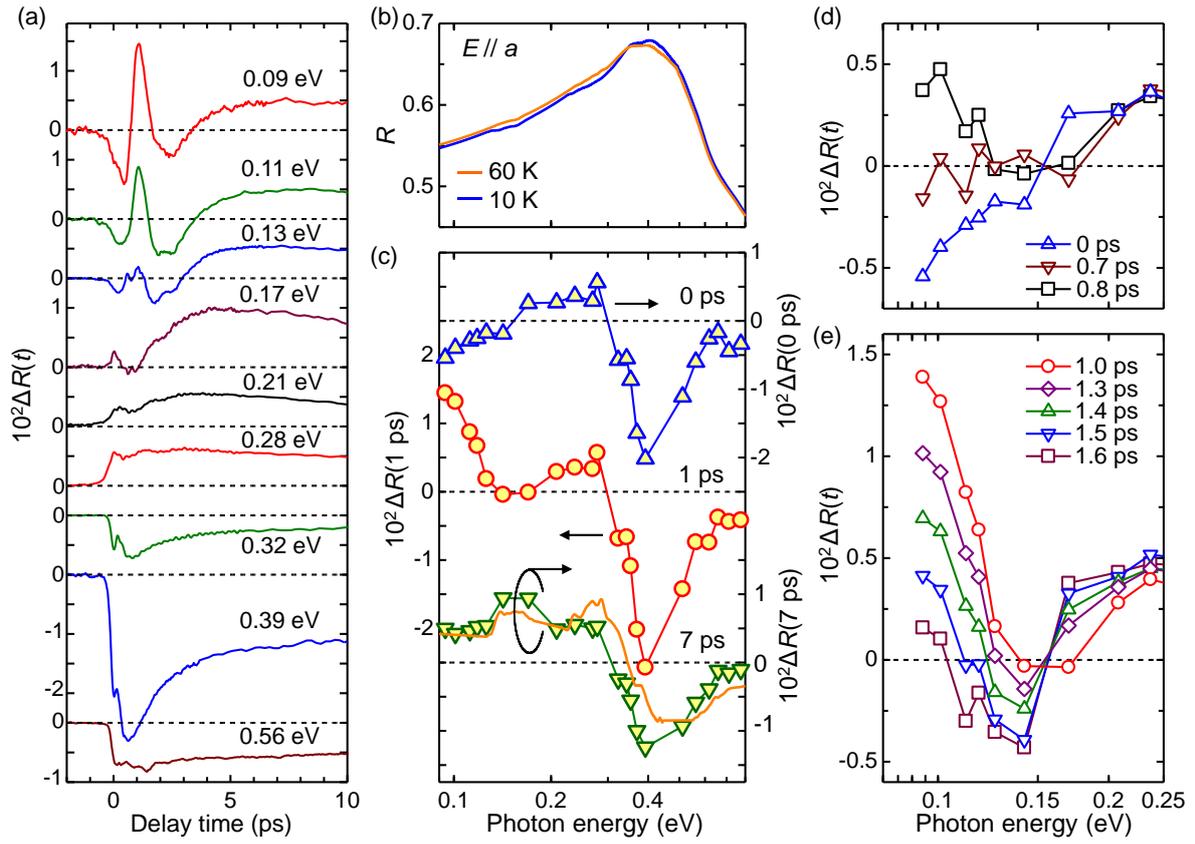

Fig. 8 Probe photon-energy dependence of reflectivity changes $\Delta R(t)$ induced by the strong terahertz pulse with $E_{\text{THz}}(0) = 1.65$ MV/cm at 10 K. (a) Time characteristics of $\Delta R(t)$ at various probe energies. (b) $R$ spectra for $E \mathbin{/\mkern-6mu/} a$ at 10 K and 60 K. (c) $\Delta R(t)$ spectra at 0 ps, 1 ps, and 7 ps. The orange line is the differential reflectivity spectrum between 60 K and 10 K, $(R_{60\text{ K}} - R_{10\text{ K}})$. (d,e) Expanded $\Delta R(t)$ spectra in the MIR region for (d) $t < 1$ ps and (e) $t \geq 1$ ps.

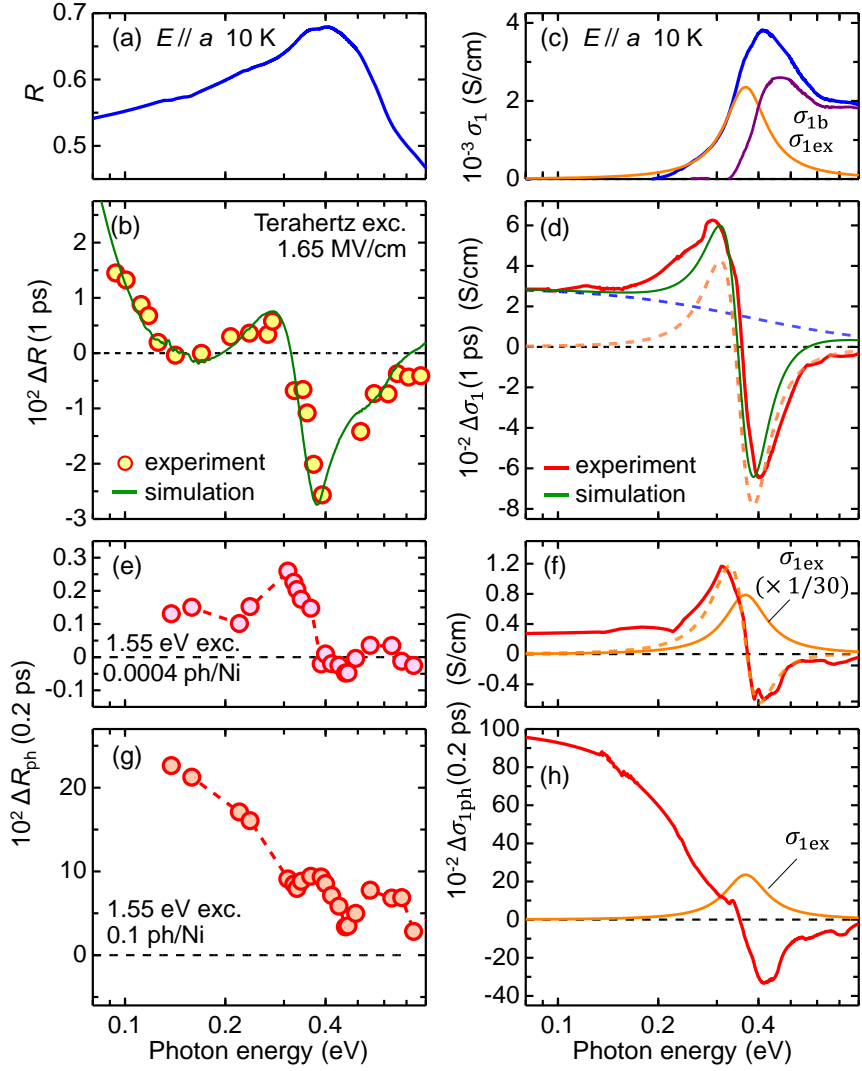

Fig. 9 Analyses of $\Delta R(t)$ and $\Delta\sigma_1(t)$ spectra at 1 ps for $E_{\text{THz}}(0) = 1.65$ MV/cm (10 K). (a) The $R$ spectrum for $E \mathbin{/\mkern-6mu/} a$ at 10 K. (b) The $\Delta R(1\text{ ps})$ spectrum (the red circles) and the simulation curve (the green line). (c) The $\sigma_1$ spectrum for $E \mathbin{/\mkern-6mu/} a$ at 10 K (the blue line) and the components of the transition specific to the exciton-condensed state, $\sigma_{1\text{ex}}$ (the orange line) and of the interband transition, $\sigma_{1\text{b}}$ (the brown line). This panel is the same as Fig. 5(c). (d) $\Delta\sigma_1(1\text{ ps})$ spectrum (the red line) and the simulation curve (the green line). The blue and orange broken lines show the Drude component and the change of the transition, $\sigma_{1\text{ex}}$ [the orange line in (c)] specific to the exciton-condensed state. (e,f) The spectra of (e) $\Delta R(t)$ (the red circles [46]) and (f) $\Delta\sigma_1(t)$ (the red line) at 0.2 ps obtained by the significantly weak 1.55-eV excitation with an excitation photon density of 0.0004 ph/Ni. The orange broken line in (f) shows the change of the transition, $\sigma_{1\text{ex}}$ [the orange line in (f)]. (g,h) The spectra of (g) $\Delta R(t)$ (the red circles [46]) and (h) $\Delta\sigma_1(t)$ (the red line) at 0.2 ps obtained by the strong 1.55-eV excitation with an excitation photon density of 0.1 ph/Ni. The orange line in (g) shows $\sigma_{1\text{ex}}$.

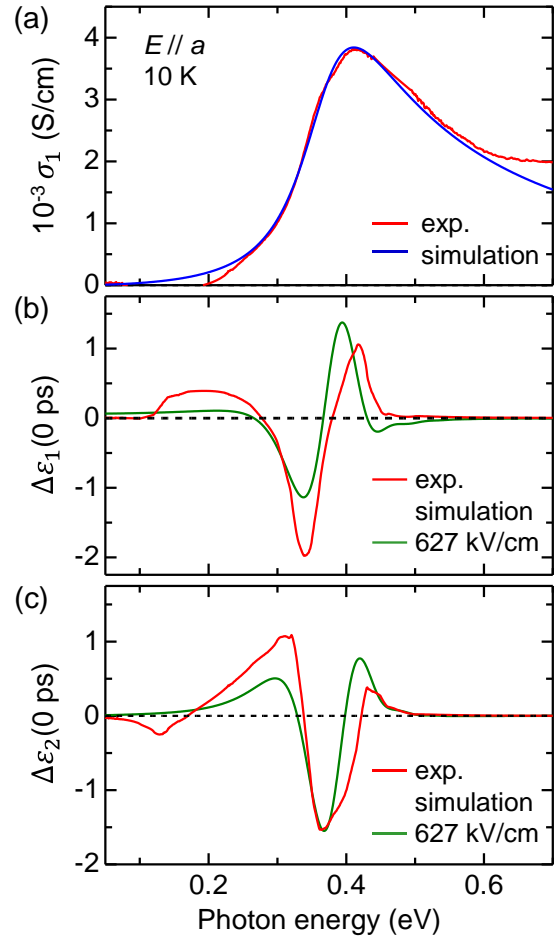

Fig. 10 Simulations of $\Delta\varepsilon_1(0\ \text{ps})$ and $\Delta\varepsilon_2(0\ \text{ps})$ spectra derived from the electro-reflectance spectrum shown in Fig. 5(b) based upon the Franz-Keldysh effect. (a) $R$ spectrum for $E\ //\ a$ at 10 K (the red line) and its fitting curve (the blue line). (b,c) (b) $\Delta\varepsilon_1(0\ \text{ps})$ and (c) $\Delta\varepsilon_2(0\ \text{ps})$ spectra (the red lines) and the simulation curves for $E_{\text{THz}}(0) = 627$ kV/cm (the green lines).

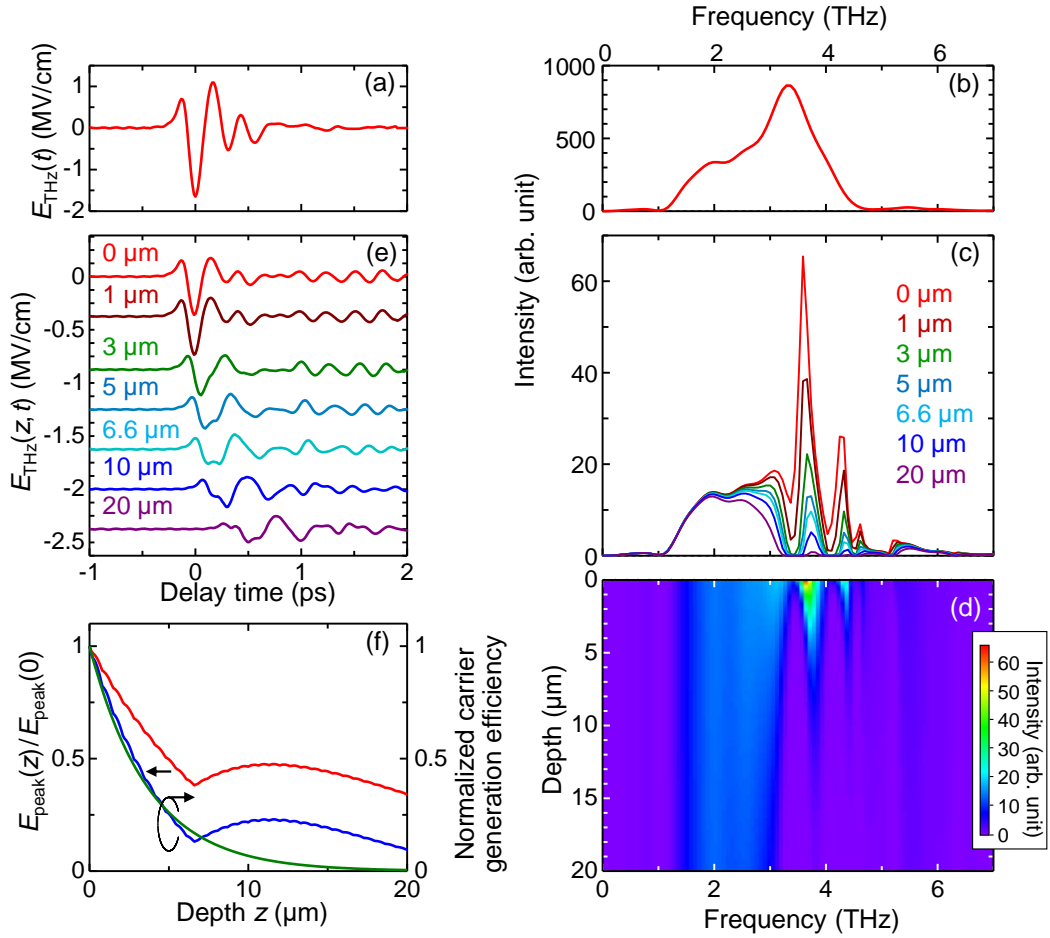

Fig. 11 Simulations of electric-field waveforms $E_{\text{THz}}(z,t)$ of a terahertz pump pulse inside the crystal. (a) Electric-field waveform of the terahertz pulse. (b) Fourier power spectrum of the terahertz pulse shown in (a). (c) Power spectra of the terahertz pulse inside the crystal for various distances from the crystal surface, $z$. (d) Contour map of Fig. (c). (e) Electric-field waveforms $E_{\text{THz}}(z,t)$ for various distances $z$. (f) Peak values of $E_{\text{THz}}(z,t)$, $E_{\text{Peak}}(z)$, normalized by $E_{\text{Peak}}(0)$ (the red line). The normalized carrier generation efficiency as a function of $z$ and the curve approximating it with a single exponential function are shown by the blue and green lines, respectively.

TABLE I  Physical parameters of excitonic states derived from the fitting of $\chi^{(3)}(-\omega; 0,0, \omega)$ spectrum with the three-level model.

| | |
|---|---|
| $\hbar\omega_1$ | 0.360 eV |
| $\hbar\gamma_1$ | 0.065 eV |
| $\langle 0|x|1\rangle$ | 1.75 Å |
| $\hbar\omega_2$ | 0.396 eV |
| $\hbar\gamma_2$ | 0.097 eV |
| $\langle 1|x|2\rangle$ | 0.95 Å |

TABLE II  Physical parameters derived from the fitting of $\Delta R$ spectrum at $t = 1$ ps shown in Fig. 9(b).

| | |
|---|---|
| $\Delta I/I$ | -0.0343 |
| $\hbar\Delta\omega_1$ | -0.028 eV |
| $\hbar\omega_p$ | 0.90 eV |
| $\hbar\gamma_D$ | 0.372 eV |
| $\Delta\varepsilon_\infty$ | 2.0 |